\documentclass[aps,pre,showpacs,amsmath,amsfonts,amssymb,superscriptaddress,twocolumn,10pt]{revtex4-1}
\usepackage{graphicx}
\usepackage{bm}
\usepackage{bbm}
\usepackage{xcolor}
\usepackage[colorlinks,linkcolor=blue,anchorcolor=blue,citecolor=blue,filecolor=blue,menucolor=blue,urlcolor=blue]{hyperref}

\newcommand{\beq}{\begin{equation}}
\newcommand{\eeq}{\end{equation}}
\newcommand{\bea}{\begin{eqnarray}}
\newcommand{\eea}{\end{eqnarray}}

\newcommand{\NN}{{\mathbbm N}}
\newcommand{\RR}{{\mathbbm R}}

\newcommand{\deemph}[1]{{\color{black!40}#1}}

\input epsf

\begin{document}

\title{Exploring the energy landscape of \texorpdfstring{$XY$}{XY} models}

\author{Rachele Nerattini}
\email{rachele.nerattini@fi.infn.it}
\affiliation{Dipartimento di Fisica e Astronomia and Centro per lo Studio
delle Dinamiche Complesse (CSDC), Universit\`a di
Firenze, via G.~Sansone 1, I-50019 Sesto Fiorentino (FI), Italy}
\affiliation{Istituto Nazionale di Fisica Nucleare (INFN), Sezione di
Firenze, via G.~Sansone 1, I-50019 Sesto Fiorentino (FI), Italy}
\author{Michael Kastner}
\email{kastner@sun.ac.za}
\affiliation{National Institute for Theoretical Physics (NITheP), Stellenbosch 7600, South Africa}
\affiliation{Institute of Theoretical Physics, University of Stellenbosch, Stellenbosch 7600, South Africa}
\author{Dhagash Mehta}
\email{dbmehta@syr.edu}
\affiliation{Department of Physics, Syracuse University, Syracuse, New York 13244, USA}
\author{Lapo Casetti}
\email{lapo.casetti@unifi.it}
\affiliation{Dipartimento di Fisica e Astronomia and Centro per lo Studio
delle Dinamiche Complesse (CSDC), Universit\`a di
Firenze, via G.~Sansone 1, I-50019 Sesto Fiorentino (FI), Italy}
\affiliation{Istituto Nazionale di Fisica Nucleare (INFN), Sezione di
Firenze, via G.~Sansone 1, I-50019 Sesto Fiorentino (FI), Italy}

\date{\today}

\begin{abstract}
We investigate the energy landscape of two- and three-dimensional $XY$ models with nearest-neighbor interactions by analytically constructing several classes of stationary points of the Hamiltonian. These classes are analyzed, in particular with respect to possible signatures of the thermodynamic phase transitions of the models. We find that, even after explicitly breaking the global O(2) symmetry of the $XY$ spins, an exponentially large class of stationary points are singular and occur in continuous one-parameter families. This property may complicate the use of theoretical tools developed for the investigation of phase transitions based on stationary points of the energy landscape, and we discuss strategies to avoid these difficulties. 
\end{abstract}


\maketitle

\section{Introduction}
\label{sec_intro}
The stationary points of a function of many variables $f : M \mapsto \mathbb{R}$ are the points $p^\text{s}\in M$ on a manifold $M$ such that $\nabla f(p^\text{s}) = 0$. Stationary points play an important role for quite a few methods in theoretical physics, as knowledge about these points can be used to infer physical properties of the system under investigation. When the function $f$ is the energy of a many-body system and the manifold $M$ is the phase space or the configuration space of the system, these methods are referred to as {\em energy landscape methods}\/ \cite{Wales:book}. Examples of applications include clusters \cite{Wales:book}, disordered systems and glasses \cite{DebenedettiStillinger:nature2001,Sciortino:jstat2005}, biomolecules, and protein folding \cite{OnuchicLutheyWolynes:arpc1997}. Based on knowledge about the stationary points of the energy function, landscape methods can be applied to estimate dynamic as well as static properties. In most applications, like for example Stillinger and Weber's thermodynamic formalism \cite{StillingerWeber:science1984,Stillinger:science1995} and other `superposition approaches' \cite{Wales:book,wales93f} for the study of equilibrium properties, only minima of the energy landscape are taken into account. In some later work, also first-order saddles (see e.g.\ Ref.\ \cite{StrodelWales:cpl2008}) and stationary points of arbitrary index \footnote{The index of a stationary point $p^\text{s} \in M$ of a function $f : M \mapsto \mathbb{R}$ is the number of unstable directions, i.e., the number of negative eigenvalues of the Hessian of $f$ at $p^\text{s}$.} have been considered, for instance to characterize glassy behavior \cite{AngelaniEtAl:prl2000,GrigeraCavagnaGiardinaParisi:prl2002}.

The potential relevance of energy landscape properties for equilibrium phase transitions was suggested after it was realized that stationary points of the Hamiltonian are related to topology changes of the phase space accessible to the system. It was conjectured that some of these topology changes, and therefore some of the stationary points, are at the origin of thermodynamic phase transitions \cite{prl1997,prl1999,physrep2000,pre2002,jsp2003}; quite some research activity followed \cite{epl2003,GaraninSchillingScala:pre2004,GrinzaMossa:prl2004,RibeiroStariolo:pre2004,AndronicoEtAl:pre2004,Kastner:prl2004,pre2005,AngelaniRuoccoZamponi:pre2005,Risau-GusmanEtAl:prl2005,HahnKastner:pre2005,jpa2006,Risau-GusmanEtAl:jsp2006,AngelaniRuocco:pre2007,AngelaniRuocco:pre2008,prerap2009,SantosCoutinho-Filho:pre2009,Kastner:pre2011,MehtaKastner:annphys2011,Mehta:2011xs,Baroni:jstat2011,FarberFromm:jams2011,CarlssonEtAl:pre2012}, some focused on specific models, others trying to shed light on the general mechanisms (see \cite{Kastner:rmp2008,Pettini:book} for reviews).

Although equilibrium phase transitions in systems with non-fluctuating particle numbers have been mainly studied within the canonical ensemble, the connection between stationary points of the Hamiltonian and equilibrium statistical properties is more transparent in a microcanonical setting \cite{Kastner06}. This can be understood by observing that, for a system with $N$ degrees of freedom, the entropy density is defined as \footnote{Throughout the paper we set Boltzmann's constant $k_B$ to unity.}
\beq
s(\varepsilon) = \frac{1}{N}\log \omega(\varepsilon),
\eeq
where $\varepsilon = E/N$ is the energy density and $\omega$ is the density of states. For a system described by continuous variables, $\omega$ can be written as
\beq
\omega(\varepsilon) = \int_{\Gamma} \delta(\mathcal{H} - N\varepsilon) \, d\Gamma = \int_{\Gamma \cap \Sigma_{\varepsilon}} \frac{d\Sigma}{\left|\nabla \mathcal{H}\right|},
\label{coarea}
\eeq
where $\Gamma$ denotes the phase space and $d\Gamma$ its volume measure, $\Sigma_{\varepsilon}$ is the hypersurface of constant energy $E = N\varepsilon$, and $d\Sigma$ stands for the $N-1$-dimensional Hausdorff measure. The rightmost integral stems from a co-area formula \cite{Federer:book}. At a stationary point $p^\text{s}$, the gradient $\nabla\mathcal{H}(p^\text{s})$ vanishes by definition and the integrand diverges, while at the same time the measure $d\Sigma$ shrinks such that $\omega$ in general remains finite for finite systems. Indeed, a more refined analysis \cite{KSS:jstat2008} shows that, although the integral on the right-hand side of \eqref{coarea} remains finite in the vicinity of a stationary point, the density of states will be nonanalytic at stationary values $\varepsilon^\text{s}:=\mathcal{H}(p^\text{s})/N$ of the energy density for any finite $N$ \footnote{Such a behavior differs from the canonical ensemble where the canonical free energy or other thermodynamic functions may develop nonanalyticities only in the thermodynamic limit $N\to\infty$ \cite{Griffiths:inDombGreen}.}. The microcanonical nonanalyticities appearing at finite $N$ are found to be in correspondence with stationary configurations; however, the `strength' of such nonanalyticities generically decreases linearly with $N$, i.e., the first $k$ derivatives of the entropy are continuous, where $k$ is $\mathcal{O}(N)$ \cite{KSS:jstat2008,jstat2009}. The usual thermodynamic quantities, like equations of state, are given by low-order derivatives of the entropy, and the observation of nonanalyticities of order $\mathcal{O}(N)$ from noisy data is therefore restricted to very small system sizes $N$. Taking this to its logical conclusion, we expect the order of the nonanalyticities to diverge in the thermodynamic limit, leading to a vanishing effect of stationary points and smooth results for the thermodynamic functions.

Superficially, this seems to challenge the conjecture that phase transitions stem from stationary points of the energy, as it suggests that finite-$N$ nonanalyticities due to stationary points are unrelated to thermodynamic phase transitions \footnote{This challenge is only apparent. In general, the order of a nonanalyticity in the infinite system is not necessarily equal to the large-$N$ limit of the order of finite system nonanalyticities, as this identification would require two limiting procedures to commute.}. And indeed, for several model systems, phase transitions have been found to occur at energies at which no stationary points of the Hamiltonian are present \cite{Fabrizio:thesis,Kastner:prl2004,GaraninSchillingScala:pre2004,AndronicoEtAl:pre2004,HahnKastner:pre2005,AngelaniRuoccoZamponi:pre2005,AngelaniRuocco:pre2007,KastnerMehta:prl2011}. On the other hand, a substantial amount of evidence (in the form of model calculations) has accumulated in favor of the conjecture that stationary points often do play a relevant role for the emergence of phase transitions, and the presence of a transition reflects prominently in properties of the stationary points \footnote{A theorem according to which topology changes, and hence stationary points of the potential energy, would be a necessary condition for phase transition in systems with short-ranged and confining interactions has been announced in \cite{FranzosiPettini:prl} and its proof has been presented in \cite{FranzosiPettini:npb}. However, a counterexample has been found and discussed in \cite{KastnerMehta:prl2011}.}. This evidence comes mostly from exactly solvable systems (often with mean-field interactions) where the connection between stationary points and thermodynamic phase transitions has been shown explicitly \cite{prl1999,pre2002,jsp2003,epl2003,pre2005,prl2006}.
Subsequently, and of direct relevance to the present work, a possible scenario of how certain finite-$N$ singularities may survive in the thermodynamic limit has been proposed in \cite{KSS:prl2007,KastnerSchnetz:prl2008}: Only those singularities related to asymptotically flat stationary points may survive in the thermodynamic limit and induce a thermodynamic phase transition. {\em Asymptotically flat}\/ here refers to stationary points whose determinant of the Hessian matrix of the potential energy $V$ vanishes in the thermodynamic limit. More precisely, for a sequence of potential energy functions $V_N:M_N\mapsto\RR$ on configuration space $M_N$, consider a sequence of stationary points $\left\{q_N^\text{s}\in M_N\right\}_{N\in\NN}$ such that $\nabla V_N(q_N^\text{s})=0$. If the limit
\beq
v_\text{c} = \lim_{N\to\infty} V_N\left(q_N^\text{s}\right)/N
\eeq
exists and
\beq
\lim_{N\to\infty}\left|\det \text{Hess}_{V_N} \left( q_N^\text{s} \right)  \right|^{1/N} = 0,
\label{KSS}
\eeq
then the stationary points may induce a singularity in the configurational entropy density $s(v)$ at $v=v_\text{c}$ in the thermodynamic limit. This indeed was verified to happen in the exactly solvable mean-field models where the connection between stationary points of the Hamiltonian and phase transitions had been previously demonstrated \cite{KastnerSchnetz:prl2008}, and also in a non-solvable toy model of a self-gravitating particles with a phase transition between a homogeneous and a collapsed phase \cite{prerap2009}. Note that several of the above results, including the determinant criterion \eqref{KSS}, are derived under certain genericity assumptions on the Hamiltonian or potential energy function. In particular, all stationary points are required to be isolated, and we will come back to this condition in Sec.\ \ref{sec_flat}.

The main purpose of the present paper is to go beyond one-dimensional or mean-field systems by performing an analysis of stationary points and their Hessian determinants for classical $XY$ spin models with nearest-neighbor ferromagnetic interactions. In particular, we consider the $XY$ model on a two-dimensional square lattice and on a three-dimensional cubic lattice. These models are among the simplest lattice spin models with short-range interactions amenable of an energy landscape approach based on stationary points of the Hamiltonian (in the even simpler Ising model such an analysis is impossible due to the discrete character of the spin variables). For the one-dimensional $XY$ model \cite{MehtaKastner:annphys2011} as well as for the mean-field (fully connected) $XY$ model \cite{jsp2003,KastnerSchnetz:prl2008} the stationary points and their relation to phase transitions have already been studied in earlier work. In both cases a fully analytical treatment of all the stationary points was feasible, as well as an analysis of flat stationary points in the sense of Eq.\ \eqref{KSS}. The results showed a clear signature in stationary-point properties of the presence of a finite-temperature phase transition in the mean-field $XY$ model; and no signature of a transition in the one-dimensional $XY$ model, in agreement with the known thermodynamic behavior of these models.

In the more interesting case of short-range interacting $XY$ models in dimensions $d\geqslant2$, it seems unlikely that finding all stationary points of the Hamiltonian is feasible, at least for lattices of reasonable sizes (for small lattices, the stationary point analysis has been done in Refs.~\cite{Mehta:2009,Mehta:2009zv,Hughes:2012hg} using a homotopy continuation method). It is, however, possible to analytically construct certain classes of stationary points of the Hamiltonian, and this will be the main topic of the present article. Two of these classes of stationary points are large in the sense that they contain an exponentially (in the lattice size $N$) growing number of points. Exponential growth with $N$ is also expected for the overall number of stationary points \cite{StillingerWeber:science1984,WalesDoye:jcp2003,Schilling:physicad2006} and in this sense we may at least hope that the classes of stationary points we have constructed are in some way relevant for the thermodynamics of the model. We evaluated the Hessian determinant at some of these stationary points and tried to identify candidate sequences of stationary points satisfying the criterion \eqref{KSS}.

After introducing the nearest-neighbor $XY$ model in Sec.\ \ref{sec_models} and deriving expressions for stationary points and Hessian matrices in Sec.\ \ref{sec_Hessian}, a first class of what we call {\em Ising stationary configurations} is constructed in Sec.\ \ref{sec_Ising}: these are stationary points of the $XY$ model where each pair of neighboring spins is either aligned or anti-aligned. Although these configurations are easy to construct, we have to resort to sampling strategies to select members of this exponentially large class, and evaluate the Hessian determinant numerically. Earlier work \cite{prl2011,jstat2012} suggests that these Ising stationary configurations might be particularly relevant for the phase transition of the model. However, unlike in the mean-field case, no signature of the phase transitions can be discerned from the data we obtained for this class of stationary points for the two-dimensional and three-dimensional models. In Sec.\ \ref{sec_poly} we construct a class of stationary points which have the character of spin waves. Since their number is small (subexponential), they are not expected to significantly influence the thermodynamic behavior of the model. In Sec.\ \ref{sec_flat}, we construct an exponentially large class of particularly interesting stationary points whose Hessian determinant is zero. Moreover, we are able to prove that, even after explicitly breaking the global O(2) symmetry of the $XY$ model, these stationary points are not isolated, but occur in continuous families. This finding has interesting consequences which will be discussed at the end of Sec.\ \ref{sec_flat}. In Sec.\ \ref{sec_field} we investigate how the presence of inhomogeneous external magnetic fields may destroy the continuous families and lead to isolated, nonsingular stationary points.  A summary of the results and concluding remarks are presented in Sec.\ \ref{sec_conclusions}.

\section{Nearest-neighbor \texorpdfstring{$XY$}{XY} models}
\label{sec_models}

A paradigmatic class of models for the study of magnetic phase transitions (the prototype of all continuous phase transitions) are classical O$(n)$ spin lattice models. We consider $d$-dimensional hypercubic lattices with periodic boundary conditions. To each lattice site $i$ an $n$-component classical spin vector $\bm S_i = (S_i^1,\ldots,S_i^n)$ of unit length is assigned. The energy of the model is characterized by the Hamiltonian
\beq\label{H}
\mathcal{H}^{(n)} = - J \sum_{\langle i,j \rangle} \bm S_i \cdot \bm S_j= - J \sum_{\langle i,j \rangle} \sum_{a = 1}^n S^a_i S^a_j,
\eeq
where the angular brackets denote a sum over all pairs of nearest-neighboring lattice sites. The exchange coupling $J$ will be assumed to be positive, resulting in ferromagnetic interactions, and without loss of generality we set $J=1$ in the following. The Hamiltonian \eqref{H} is globally invariant under the O$(n)$ group. For the case $n=1$ the symmetry group O$(1) \equiv \mathbb{Z}_2$ is a discrete group and the Hamiltonian \eqref{H} becomes the Ising Hamiltonian
\beq\label{H_1}
\mathcal{H}^{(1)} = - \sum_{\langle i,j \rangle} \sigma_i \sigma_j,
\eeq
where $\sigma_i\in\{-1,+1\}$ $\forall i$. In all other cases $n \geqslant 2$ the O$(n)$ group is continuous. As special cases, the O$(n)$ models include the $XY$ model for $n=2$ and the Heisenberg model for $n=3$. We will consider in the following the $XY$ model on a two-dimensional square lattice and on a three-dimensional cubic lattice, hoping that this choice yields the simplest possible O$(n)$ models amenable to the energy landscape analysis we have in mind. The $XY$ spins live on the unit circle $\mathbb{S}^1$, so that the components of the $i$th spin can be parametrized by a single angular variable $\vartheta_i \in [0, 2\pi)$,
\beq\label{XY_components}
\begin{cases}
S^1_i & = \cos\vartheta_i,\\
S^2_i & = \sin\vartheta_i.
\end{cases}
\eeq
We can thus conveniently write the Hamiltonian of the $XY$ model as
\beq\label{H_XY}
\mathcal{H}^{(2)} = - \frac{1}{2}\sum_{i = 1}^N \sum_{j\in\mathcal{N}(i)} \cos\left(\vartheta_i - \vartheta_j \right),
\eeq
where $\mathcal{N}(i)$ denotes the set of nearest neighbors of lattice site $i$. The energy density $\varepsilon = \mathcal{H}^{(2)}/N$ takes values in the interval $[-d,d]$ where $d$ is the lattice dimension.

In two dimensions, the $XY$ model exhibits a Bere\-\v{z}inskij-Kosterlitz-Thouless (BKT) phase transition \cite{berezinskij:sovphysjetp1971,KosterlitzThouless:jphysc1973} with no long-range order, while in three dimensions it undergoes a ferromagnetic transition in the same universality class as that of the superfluid transition \cite{Goldenfeld:book}. With our choices for the lattices and the interaction constants, the BKT transition in the two-dimensional case occurs at a critical temperature $T^{2d}_\text{c} \approx 0.8929(1)$ corresponding to a critical energy density $\varepsilon^{2d}_\text{c} = \langle \mathcal{H}^{(2)}\rangle\left(T_\text{c}\right)/N \approx -1.4457(4)$ \cite{GuptaBaillie:prb1992,Hasenbusch:jphysa2005}, while the critical temperature of the ferromagnetic transition in three dimensions is $T^{3d}_\text{c} \approx 2.20167(9)$, corresponding to a critical energy density $\varepsilon^{3d}_\text{c} \approx -0.99184(6)$ \cite{GottlobHasenbusch:physicaa1993}.

\section{Stationary points and Hessian of the energy}
\label{sec_Hessian}

To conduct an energy landscape analysis as outlined in the Introduction, we have to find the stationary points of the energy, i.e., solutions $\vartheta^\text{s}$ satisfying the vector equation $\nabla \mathcal{H}^{(2)}(\vartheta^\text{s}) = 0$. Using \eqref{H_XY}, the $k$th component of this equation can be written as
\beq\label{stationary_points}
\sum_{j\in\mathcal{N}(k)} \sin \left(\vartheta_k - \vartheta_j \right) = 0.
\eeq
The O$(2)$ invariance of the Hamiltonian \eqref{H_XY} implies that the solutions of \eqref{stationary_points} are not isolated points in configuration space, but occur in continuous curves: Given a stationary point $\vartheta^\text{s}=(\vartheta^\text{s}_1,\dotsc,\vartheta^\text{s}_N)$, the points $(\vartheta^\text{s}_1+\alpha,\dotsc,\vartheta^\text{s}_N+\alpha)$ are also stationary points for arbitrary $\alpha\in\RR$. Several of the theoretical tools and results mentioned in the Introduction, and in particular the Hessian determinant criterion \eqref{KSS}, require energy functions with only isolated stationary points. It is therefore necessary to explicitely break the global O$(2)$ symmetry of the $XY$ model. Here we choose to fix one spin, e.g.\ $\vartheta_N \equiv 0$, but there are other possibilities how the symmetry breaking can be achieved, and we will come back to this point in Sec.\ \ref{sec_field}. In the large-$N$ limit, the only effect of this global phase fixing is to dictate the direction of the spontaneous symmetry breaking, but otherwise thermodynamic quantities remain unaffected.

In order to apply the criterion \eqref{KSS}, we have to evaluate, at the stationary points, the determinant of the Hessian matrix of the Hamiltonian [which for the $XY$ model \eqref{H_XY} coincides with the potential energy]. The elements of the Hessian matrix are defined as
\beq
H_{kl} = \frac{\partial^2 \mathcal{H}}{\partial \vartheta_k \partial \vartheta_l}.
\eeq
The constraint $\vartheta_N \equiv 0$ makes the Hessian an $(N-1)\times(N-1)$ matrix and, for the $XY$ Hamiltonian \eqref{H_XY}, its diagonal elements are given by
\beq\label{H_diag}
H_{kk} = \sum_{j\in\mathcal{N}(k)} \cos \left(\vartheta_k - \vartheta_j \right),
\eeq
while the off-diagonal elements are
\beq\label{H_offdiag}
H_{kl} = \begin{cases}
- \cos \left(\vartheta_k - \vartheta_l \right) & \text{for $l\in\mathcal{N}(k)$},\\
0 & \text{else},
\end{cases}
\eeq
for $k,l = 1,\dotsc,N-1$.

\section{Ising stationary configurations}
\label{sec_Ising}
Finding {\em all}\/ stationary points of the Hamiltonian \eqref{H_XY} is unlikely to be feasible for large lattices. This section and the subsequent Secs.\ \ref{sec_poly} and \ref{sec_flat} will be concerned with the construction of special classes of stationary points and with the evaluation of their Hessian determinant.

Inspection of the stationary point conditions \eqref{stationary_points} reveals that any configuration where $\vartheta_i^\text{s} = \{0,\pi\}$ $\forall i$ is a stationary point, as in this case each term of the sum on the left-hand side of \eqref{stationary_points} vanishes separately. In the notation of \eqref{XY_components} such stationary points can be written as
\beq\label{e:Ising}
\begin{cases}
S^1_i & = \sigma_i, \\
S^2_i & = 0,
\end{cases}
\eeq
where $\sigma_i\in\{-1,+1\}$. Therefore, as already discussed in \cite{prl2011}, each such stationary point $\vartheta^\text{s}$ of the $XY$ Hamiltonian \eqref{H} corresponds to a configuration of the Ising model \eqref{H_1} defined on the same lattice. Moreover, the corresponding stationary values $\mathcal{H}^{(2)}(\vartheta^\text{s})$ of these `Ising stationary configurations' are just the energy levels of the Ising Hamiltonian \eqref{H_1}.

As we will see in later sections, the Ising stationary configurations are not the only stationary points of the $XY$ model, but they constitute a large class in the sense that their number grows like $2^N$ with the system size $N$. But, as already mentioned, also the overall number of stationary points of a generic function of $N$ variables is expected to scale exponentially with $N$, a fact that may be seen as a hint towards the relevance of the Ising stationary configurations. Moreover, as discussed in \cite{prl2011,jstat2012}, the critical energy densities of O$(n)$ models are remarkably close to those of the corresponding Ising model, another indication pointing towards the relevance of the Ising configurations for the phase transition of the $XY$ model.

Evaluated at the Ising stationary configurations, the Hessian matrix elements \eqref{H_diag} and \eqref{H_offdiag} can be written as
\beq\label{H_ising_diag}
H_{kk} = \sigma_k\sum_{j\in\mathcal{N}(k)} \sigma_j
\eeq
and
\beq\label{H_ising_offdiag}
H_{kl} = \begin{cases}
- \sigma_k\sigma_l & \text{for $l\in\mathcal{N}(k)$},\\
0 & \text{else},
\end{cases}
\eeq
for $k,l = 1,\ldots,N-1$ and $k\neq l$. In the mean-field $XY$ model \cite{jsp2003}, and also in other mean-field or one-dimensional models \cite{prerap2009,KastnerSchnetz:prl2008,MehtaKastner:annphys2011}, the energy and the Hessian determinant of Ising stationary configurations depend on only a single collective variable, thus allowing an analytical search of stationary points satisfying \eqref{KSS}. Unfortunately, for the two- and three-dimensional nearest-neighbor models this is not the case and we have to resort to numerical methods. We computed the determinant of the Hessian of the Hamiltonian on a numerically obtained sample of the Ising stationary configurations. The sample was obtained by standard Metropolis Monte Carlo simulations of the two-dimensional and three-dimensional Ising models, exploiting the above mentioned one-to-one relation between configurations of the Ising model and Ising stationary configurations of the $XY$ models.

\subsection{Two-dimensional \texorpdfstring{$XY$}{XY} model}
\label{sec_Ising_A}

\begin{figure}
\centering
\includegraphics[scale=3]{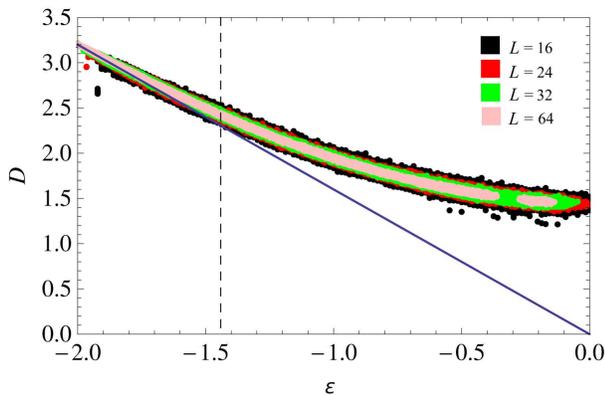}
\caption{(Color online) Rescaled Hessian determinant $D$ of Ising stationary configurations for the two-dimensional $XY$ model, plotted as a function of the energy density $\varepsilon$. Data symbols correspond to lattices of side lengths $L$ ranging from 16 to 64 (see the legend in the plot for the color/grayscale code). The critical energy density $\varepsilon^{2d}_\text{c} \approx -1.446$ of the BKT transition is marked by a vertical dashed line. The solid lines are the values calculated for the polygonal configurations in the large-$N$ limit according to \eqref{normdetpol2d_eps}.}
\label{fig_ising_xy2d}
\end{figure}

We considered $L\times L$ square lattices of side lengths $L = 16$, 24, 32 and 64, so that the number of degrees of freedom ranges from $N=L^2=256$ to 4096. Compared to those typically considered in simulations nowadays, these are not very big lattices, and indeed obtaining the sample was easy and fast. The practical limit on the number of degrees of freedom was set by the time-consuming calculation of the Hessian determinant for each configuration of the sample. Although in principle Ising configurations occur over the entire range $[-2,2]$ of accessible energy densities, only configurations with negative energy were sampled in the Monte Carlo runs. This is a consequence of using canonical simulations at positive simulation temperature so that, for sufficiently large lattice sizes, the Boltzmann weight narrowly focuses the sampled distribution on a range of negative energies. However, by using also negative temperatures we would have obtained symmetric results with respect to zero energy, without adding any relevant information. For each lattice of side lengths $L=16$, 24 and 32, we considered a total sample of $250000$ configurations. For $L=64$ we considered only $48000$ configurations, the Hessian determinant being quite heavy to compute. Results for the rescaled Hessian determinant
\begin{equation}\label{e:D}
D=\left|\det \text{Hess}_{\mathcal{H}^{(2)}} \left( \vartheta^\text{s} \right)  \right|^{1/N}
\end{equation}
as a function of the energy density are shown in Fig.\ \ref{fig_ising_xy2d}.

\begin{figure}
\centering
\includegraphics[scale=0.2]{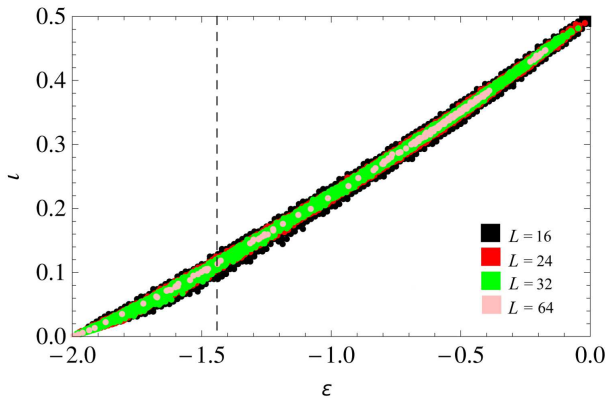}
\caption{(Color online) As in Fig. \ref{fig_ising_xy2d} for the index density $\iota$. Data for the polygonal configurations are not shown.}
\label{index_2d}
\end{figure}

In order to further characterize the sampled stationary points we computed, for the same lattices, the index density
\beq
\iota = \frac{\text{index}(\vartheta^\text{s})}{N-1},
\label{ind_dens}
\eeq
where the index of a stationary point $\vartheta^\text{s}$ is the number of negative eigenvalues of the Hesse matrix at $\vartheta^\text{s}$. 
The results for the index density $\iota$ as a function of the energy density are shown in Fig.\ \ref{index_2d}.


Two features of the results shown in Figs.\ \ref{fig_ising_xy2d} and \ref{index_2d} are of particular interest.
\begin{enumerate}
\item As $N$ grows, the rescaled determinant $D$ as well as the index density $\iota$ show a tendency to concentrate onto a single curve, so that, at least for Ising stationary configurations, these quantities appear to be good thermodynamic observables. Moreover, both quantities appear to be monotonic functions of the energy density.
\item The Hessian determinant shows no tendency to vanish for any value of the energy density. Hence there are no indications of the presence of asymptotically flat stationary points, i.e., of the validity of Eq.\ \eqref{KSS} around the transition energy density $\varepsilon^{2d}_\text{c} \approx -1.446$ of the BKT transition. Also the index density $\iota(\varepsilon)$ does not show any remarkable feature close to $\varepsilon^{2d}_\text{c}$.
\end{enumerate}

Our sample has variable magnetization and, in particular for low energies, configurations typically have nonzero magnetizations while in the two-dimensional $XY$ model the typical magnetization is zero at any energy. In order to rule out the possibility that this may affect our results, we repeated the calculation of the Hessian determinant on a sample of configurations with vanishing magnetization, obtained by Monte Carlo with Kawasaki dynamics. The results (not shown) display no appreciable differences with respect to Fig.\ \ref{fig_ising_xy2d}. 

\subsection{Three-dimensional \texorpdfstring{$XY$}{XY} model}
\label{sec_Ising_B}

\begin{figure}
\centering
\includegraphics[scale=0.2]{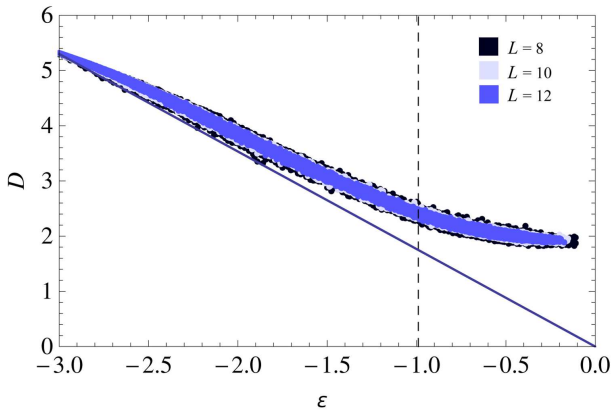}
\caption{(Color online) Rescaled Hessian determinant $D$ as a function of the energy density $\varepsilon$ for the three-dimensional $XY$ model. Data symbols correspond to lattices of side lengths $L$ ranging from 8 to 12 (see the legend in the plot for the color/grayscale code). The critical energy density $\varepsilon^{3d}_\text{c} \approx -0.99$ of the ferromagnetic transition is marked by a vertical dashed line. The solid lines are the values calculated for the polygonal configurations in the large-$N$ limit according to \eqref{normdetpol3d_eps}.}
\label{fig_ising_xy3d}
\end{figure}

\begin{figure}
\centering
\includegraphics[scale=0.2]{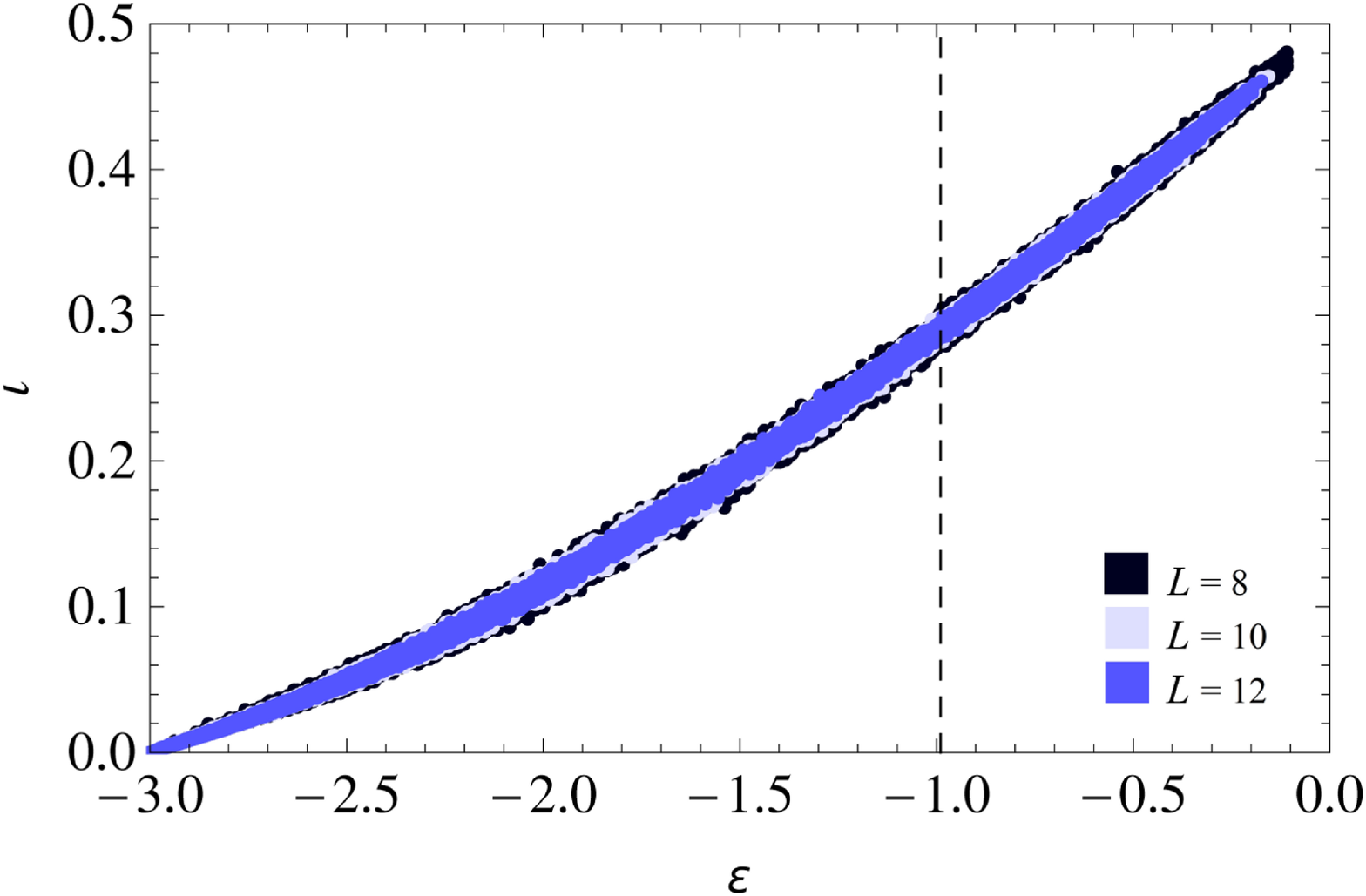}
\caption{(Color online) As in Fig. \ref{fig_ising_xy3d} for the density of index $\iota$. Data for the polygonal configurations are not shown.}
\label{index_3d}
\end{figure}

In the three-dimensional case we proceeded analogously to the two-dimensional case, considering $L\times L\times L$ lattices of side lengths $L=8$, 10 and 12, so that the number of degrees of freedom ranged from $N=L^3=512$ to 1728. For each lattice we considered a total sample of $57000$ configurations. Results for the rescaled Hessian determinant $D$ and for the index density $\iota$ as a function of the energy density are shown in Figs.\ \ref{fig_ising_xy3d} and \ref{index_3d}. The similarities to Figs.\ \ref{fig_ising_xy2d} and \ref{index_2d} are striking, so that the considerations made for the two-dimensional case carry over to three dimensions.

\section{Polygonal stationary points}
\label{sec_poly}

Another class of stationary configurations of the $XY$ model that can be easily identified are those for which neighboring spins differ by the same angle $\varphi$ \footnote{These configurations can be generalized to the case in which there is a different constant angle for each of the $d$ independent directions of the lattice; however, for simplicity we shall restrict to the case of just one angle, equal for all the directions.},
\beq\label{poly}
\vartheta_j = \vartheta_i \pm \varphi \qquad \forall i = 1,\ldots,N,\;j\in\mathcal{N}(i).
\eeq
Periodic boundary conditions restrict these angles to values $\varphi = 2\pi m/L$ with $m\in\{0,\ldots,L-1\}$. The stationary configurations \eqref{poly} are analogous to those called `spinwave' stationary points for the 1-$d$ $XY$ model in \cite{MehtaKastner:annphys2011} and to the `polygonal' stationary points of the self-gravitating ring model introduced in \cite{prerap2009}. We will adopt the latter naming convention. 

For the two-dimensional $XY$ model, the energy density of a polygonal stationary configuration is
\beq
\varepsilon(\varphi) = -2 \cos\varphi,
\label{energy_pol2d}
\eeq
and the Hessian determinant of the Hamiltonian has the simple form
\beq\label{hesspol2d}
H_{ij}(\varphi) = A_{i,j} \cos\varphi,
\eeq
where $A$ is an $(N-1) \times (N-1)$ matrix with elements
\beq\label{e:A}
A_{ij}=\begin{cases}
\hphantom{-}4 & \text{if $i=j$},\\
-1 & \text{if $j\in\mathcal{N}(i)$},\\
\hphantom{-}0 & \text{else}.
\end{cases}
\eeq
\begin{figure}
\centering
\includegraphics[scale=1.05]{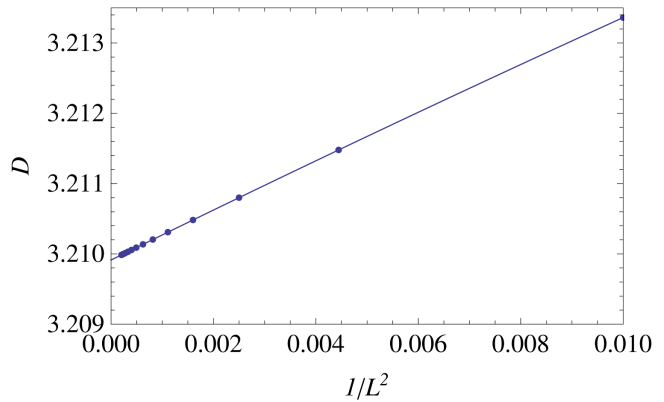}
\caption{(Color online) Rescaled determinant $D=|\det A|^{1/(N-1)}$ of the matrix $A$ as defined in \eqref{e:A}, plotted as a function of the inverse system size. The line is obtained from a linear least-square fit, and an extrapolation to $1/L^2=0$ yields $a=\lim_{N\to\infty}D\approx3.21$.}
\label{f:polig}
\end{figure}%
We have analyzed the determinant of $A$ numerically, and the results shown in Fig.\ \ref{f:polig} provide strong evidence that, asymptotically for large $N$, the determinant behaves as
\beq\label{detA}
\det A \sim a^{N-1}
\eeq
with $a\approx3.21$. The rescaled Hessian determinant computed on these configurations in the thermodynamic limit is then given by
\beq\label{normdetpol2d}
\lim_{N\to\infty} \left|\det H(\varphi) \right|^{1/(N-1)} = a|\cos\varphi|
\eeq
and, using \eqref{energy_pol2d}, we can write
\beq\label{normdetpol2d_eps}
\lim_{N\to\infty} \left|\det H(\varphi) \right|^{1/(N-1)} =\frac{a}{2}|- \varepsilon|.
\eeq
This result is plotted in Fig.\ \ref{fig_ising_xy2d} along with the data for the Ising stationary points.

For the polygonal stationary points of the three-dimensional $XY$ model, the calculation proceeds along very similar lines, yielding as a final result
\beq
\lim_{N\to\infty} \left|\det H(\varphi) \right|^{1/(N-1)} =\frac{b}{3}|- \varepsilon|
\label{normdetpol3d_eps}
\eeq
with $b\approx 5.3$. This result is plotted in Fig.\ \ref{fig_ising_xy3d}, along with data for the Ising stationary configurations.

\section{Singular stationary points}
\label{sec_flat}

In the Introduction we elaborated, amongst other things, on the criterion \eqref{KSS}, which relates the occurrence of a phase transition in the thermodynamic limit to the existence of stationary points whose Hessian determinant, in this limit, approaches zero in a suitable way. This, and other results mentioned, were derived under the assumption that the Hamiltonian $\mathcal{H}$ of the system under consideration is a Morse function, meaning that at any stationary point of $\mathcal{H}$ the Hessian determinant is nonzero. In principle, this requirement is not particularly restrictive: Morse functions are known to be generic, in the sense that, even if a given function is not a Morse function, it can be made into one by adding a generic perturbation (see Chapter 4.4 and also the introductory comments in Chapter 3.1 of \cite{Demazure}). Although this property underlines the relevance of Morse function for realistic (and therefore imperfect) physical systems, the Morse property may well be violated in clean, idealized physical models. To avoid this problem, one could of course add a small generic perturbation to the model, but this will typically make analytic calculations much harder, if not impossible.

In the following we prove that, in lattice dimensions $d=2$ and $d=3$, the $XY$ Hamiltonian \eqref{H_XY} is not a Morse function, but instead has an exponentially (in $N$) large number of singular stationary points. Moreover, the stationary energy densities $\mathcal{H}(\vartheta^\text{s})/N$ of all these singular stationary points become dense on the interval $[-d,d]$ of accessible energy densities in the thermodynamic limit. The proof is constructive, and for simplicity we restrict the presentation to two-dimensional square lattices of size $L\times L$ with periodic boundary conditions. The three-dimensional case is treated in Appendix \ref{s:A}. Generalizations to higher-dimensional lattices should be possible along similar lines, but we did not work this out in detail.

For a configuration $\vartheta=(\vartheta_1,\dotsc,\vartheta_N)$ to have a vanishing Hessian determinant, it is sufficient that one row of the Hessian matrix given in \eqref{H_diag} and \eqref{H_offdiag} has only zero entries. This is a local property, as all the nonzero entries in the $k$th row are fully determined by the $k$th spin and its nearest neighbors. Consider for example a configuration which, somewhere on the lattice, locally looks like
\begin{equation}\label{e:local_config}
\begin{matrix}
\cdot & \downarrow & \cdot\\
\uparrow & \leftarrow & \uparrow\\
\cdot & \downarrow & \cdot
\end{matrix}
\end{equation}
where arrows $\uparrow$, $\rightarrow$, $\downarrow$, $\leftarrow$ correspond to angle variables $\vartheta_i=0$, $\pi/2$, $\pi$, $3\pi/2$. The dots in \eqref{e:local_config} are place holders for arbitrary spin orientations, as their values do not matter for the moment. Assigning to the center (left-pointing) spin of this configuration the label $k$, we find
\begin{equation}
H_{kk}=\sum_{j\in\mathcal{N}(k)}\cos(\pm\pi/2)=0,\qquad H_{kl}=0\;\forall l
\end{equation}
for the elements of the Hessian matrix. The matrix therefore does not have full rank and its determinant is zero.

Next, in addition to this local condition guaranteeing that the Hessian determinant vanishes, we also have to ensure that the overall configuration is a stationary point of $\mathcal{H}^{(2)}$. This is a global property as, in order for a configuration to be stationary, the constraint
\begin{equation}\label{e:stationarity}
\sum_{l\in\mathcal{N}(k)} \sin\left(\vartheta_k - \vartheta_l\right)=0
\end{equation}
has to be satisfied for all lattice sites $k$. Starting from the pattern in \eqref{e:local_config}, it is not too difficult to construct an embedding of such patterns into larger lattices while at the same time satisfying the stationarity constraints \eqref{e:stationarity}. For the example of an $8\times8$ square lattice, the following class of configurations does the job,
\begin{equation}\label{e:ssp}
\renewcommand{\arraystretch}{1.37}
\begin{matrix}
\uparrow & \deemph\updownarrow & \deemph\updownarrow & \deemph\updownarrow & \deemph\updownarrow & \deemph\updownarrow & \uparrow & \leftarrow\\
\deemph\updownarrow & \deemph\updownarrow & \deemph\updownarrow & \deemph\updownarrow & \deemph\updownarrow & \downarrow & \rightarrow & \downarrow\\
\deemph\updownarrow & \deemph\updownarrow & \deemph\updownarrow & \deemph\updownarrow & \uparrow & \leftarrow & \uparrow & \deemph\updownarrow\\
\deemph\updownarrow & \deemph\updownarrow & \deemph\updownarrow & \downarrow & \rightarrow & \downarrow & \deemph\updownarrow & \deemph\updownarrow\\
\deemph\updownarrow & \deemph\updownarrow & \uparrow & \leftarrow & \uparrow & \deemph\updownarrow & \deemph\updownarrow & \deemph\updownarrow\\
\deemph\updownarrow & \downarrow & \rightarrow & \downarrow & \deemph\updownarrow & \deemph\updownarrow & \deemph\updownarrow & \deemph\updownarrow\\
\uparrow & \leftarrow & \uparrow & \deemph\updownarrow & \deemph\updownarrow & \deemph\updownarrow & \deemph\updownarrow & \deemph\updownarrow\\
\rightarrow & \downarrow & \deemph\updownarrow & \deemph\updownarrow & \deemph\updownarrow & \deemph\updownarrow & \deemph\updownarrow & \downarrow
\end{matrix}
\end{equation}
The lattice sites marked with gray $\updownarrow$-arrows can be filled with an arbitrary `Ising-type'-pattern of $\uparrow$ and $\downarrow$ arrows. Independently of the precise pattern of these up- and down-pointing arrows, the resulting configuration will always be stationary. In this way, we have obtained a class of stationary points of the Hamiltonian $\mathcal{H}^{(2)}$ with vanishing Hessian determinant, and the scheme works in just the same way for larger lattice sizes.

This class of singular stationary points $\vartheta^\text{s}$ is ample enough to allow us to adjust the energy density $\mathcal{H}(\vartheta^\text{s})/N$ almost freely: By choosing an appropriate Ising-type pattern of $\uparrow$ and $\downarrow$ for the gray $\updownarrow$-arrows in \eqref{e:ssp}, the energy of the configuration is varied. Since the number of gray $\updownarrow$-arrows in such a configuration scales as $L^2$, their contribution to the overall energy will, in the large-$L$ limit, dominate over the fixed (black) arrows in \eqref{e:ssp} whose number increases only linearly in $L$. As a result, the corresponding stationary energy densities $\mathcal{H}(\vartheta^\text{s})/N$ are dominated by the Ising-type pattern chosen for the gray $\updownarrow$-arrows and, like the Ising energy densities, become dense on the interval $[-2,2]$ of accessible energy densities in the thermodynamic limit.

Singular stationary points come in two flavors: They can either be isolated stationary points, like at the minimum $x^\text{s}=0$ of the quartic $f_1(x)=x^4$. Or they can form continuous families of non-isolated stationary points, like for the Mexican hat potential $f_2(x,y)=(x^2+y^2)^2- (x^2+y^2)$ where the points on the circle $x^2+y^2=1/2$ form a continuous curve of minima of $f_2$. Our singular stationary points of the two-dimensional $XY$ Hamiltonian fall into the latter category. This can be seen by starting from a configuration like the one depicted in \eqref{e:ssp} and then simultaneously rotating by some arbitrary angle $\alpha$ all the $\rightarrow$ and $\leftarrow$ spins situated on the diagonal. It is easily checked that the resulting configuration still satisfies the stationarity condition \eqref{e:stationarity}. This proves that the singular stationary points we have constructed are not isolated, but occur in continuous one-parameter families, parametrized by the angle $\alpha$. Similarly, one can create two- and more-parameter families by generalizing \eqref{e:ssp} to contain more than one diagonal pattern,
\begin{equation}\label{e:ssp2}
\renewcommand{\arraystretch}{1.37}
\begin{matrix}
\uparrow & \deemph\updownarrow & \uparrow & \leftarrow & \uparrow & \deemph\updownarrow & \uparrow & \leftarrow\\
\deemph\updownarrow & \downarrow & \rightarrow & \downarrow & \deemph\updownarrow & \downarrow & \rightarrow & \downarrow\\
\uparrow & \leftarrow & \uparrow & \deemph\updownarrow & \uparrow & \leftarrow & \uparrow & \deemph\updownarrow\\
\rightarrow & \downarrow & \deemph\updownarrow & \downarrow & \rightarrow & \downarrow & \deemph\updownarrow & \downarrow\\
\uparrow & \deemph\updownarrow & \uparrow & \leftarrow & \uparrow & \deemph\updownarrow & \uparrow & \leftarrow\\
\deemph\updownarrow & \downarrow & \rightarrow & \downarrow & \deemph\updownarrow & \downarrow & \rightarrow & \downarrow\\
\uparrow & \leftarrow & \uparrow & \deemph\updownarrow & \uparrow & \leftarrow & \uparrow & \deemph\updownarrow\\
\rightarrow & \downarrow & \deemph\updownarrow & \downarrow & \rightarrow & \downarrow & \deemph\updownarrow & \downarrow
\end{matrix}
\end{equation}
In this configuration the $\rightarrow$ and $\leftarrow$ spins situated on the main diagonal can be simultaneously rotated by some angle $\alpha$, and those on the other diagonal (modulo periodic boundary conditions) by an independent angle $\beta$, resulting in a continuous two-parameter family of stationary points. The generalization to more parameters is straightforward, provided the lattice sizes are chosen large enough.

Note that this occurrence of continuous families of non-isolated stationary points is not due to the global $O(2)$ invariance of the $XY$ Hamiltonian: This global symmetry is a trivial effect that we have taken care of by fixing one angle variable, $\vartheta_N=0$, as discussed in Sec.\ \ref{sec_Hessian}. From the examples \eqref{e:ssp} and \eqref{e:ssp2}, however, we have learned that this global phase fixing is not sufficient to ensure that the $XY$ Hamiltonian is a Morse function with only isolated stationary points. The problem seems to be that certain spin environments, like the pattern
\begin{equation}
\renewcommand{\arraystretch}{1.37}
\begin{matrix}
\uparrow & \cdot & \cdot & \cdot & \cdot & \cdot & \uparrow & \cdot\\
\cdot & \cdot & \cdot & \cdot & \cdot & \downarrow & \cdot & \downarrow\\
\cdot & \cdot & \cdot & \cdot & \uparrow & \cdot & \uparrow & \cdot\\
\cdot & \cdot & \cdot & \downarrow & \cdot & \downarrow & \cdot & \cdot\\
\cdot & \cdot & \uparrow & \cdot & \uparrow & \cdot & \cdot & \cdot\\
\cdot & \downarrow & \cdot & \downarrow & \cdot & \cdot & \cdot & \cdot\\
\uparrow & \cdot & \uparrow & \cdot & \cdot & \cdot & \cdot & \cdot\\
\cdot & \downarrow & \cdot & \cdot & \cdot & \cdot & \cdot & \downarrow\\[-4mm]
\phantom{\leftarrow} & \phantom{\leftarrow} & \phantom{\leftarrow} & \phantom{\leftarrow} & \phantom{\leftarrow} & \phantom{\leftarrow} & \phantom{\leftarrow} & \phantom{\leftarrow}
\end{matrix}
\end{equation}
in \eqref{e:ssp}, can build a 'cage' around a lattice region such that the overall phase of the enclosed region [the diagonal in the case of \eqref{e:ssp}] is shielded from the rest of the configuration. As a consequence, breaking of the {\em global}\/ O(2) invariance of the $XY$ Hamiltonian by {\em locally}\/ fixing $\vartheta_N=0$ is not sufficient. Another way to eliminate the global $O(2)$ invariance is to use antiperiodic boundary conditions in all the $d$-directions, as proposed in Ref.\ \cite{vonSmekal:2007ns}. However, we have verified numerically that even using antiperiodic boundary conditions, isolated singular solutions as well as continuous one- and more-parameter families of solutions exist.

\section{Symmetry breaking fields}
\label{sec_field}
The observation of local versus global properties also suggests how the problem of non-isolated, singular stationary points might be solved: As mentioned at the beginning of Sec.\ \ref{sec_flat}, perturbations like
\beq\label{lin_pert}
\mathcal{H}^{(2)} = - \frac{1}{2}\sum_{i = 1}^N \sum_{j\in\mathcal{N}(i)} \cos\left(\vartheta_i - \vartheta_j \right)-\sum_{i = 1}^N h_i\vartheta_i
\eeq
and maybe also
\beq\label{cos_pert}
\mathcal{H}^{(2)} = - \frac{1}{2}\sum_{i = 1}^N \sum_{j\in\mathcal{N}(i)} \cos\left(\vartheta_i - \vartheta_j \right)-\sum_{i = 1}^N h_i\cos\vartheta_i,
\eeq
for generic values of $(h_1,\dotsc,h_N)\in\RR^N$, should ensure that the Hamiltonian has only isolated and nondegenerate stationary points, but other forms of perturbations might do the job as well. For $3\times 3$ lattices we have checked numerically that, up to numerical accuracy, the perturbations in \eqref{lin_pert} and \eqref{cos_pert} indeed destroy all singular stationary points of $\mathcal{H}^{(2)}$: Firstly, we used the numerical polynomial homotopy continuation method (Bertini software package \cite{BHSW06}) which finds all the solutions of a system of multivariate polynomial equations, including isolated singular solutions \cite{79:allgower}. This method has been recently used to study the potential energy landscape in various areas of physics \cite{Mehta:2009,Mehta:2009zv,Mehta:2011xs,Mehta:2011wj,Maniatis:2012ex,Hughes:2012hg}. We studied at least $10$ generic sets of $(h_1,\dotsc,h_N)\in\RR^N$ for both types of perturbations and verified that no isolated singular solution occur for these perturbed systems. We then used an extension of the numerical polynomial homotopy continuation method, called numerical algebraic geometry \cite{SVW:96,Mehta:2012wk}, which can find solution curves of a system of polynomial equations, combined with the method described in Ref.\ \cite{Lu06findingall}, and concluded that there is no continuous solution curve for any of the systems in the presence of a generic perturbation.

From a physical point of view, the cosine-perturbed Hamiltonian \eqref{cos_pert} appears particularly appealing as it has the form of a spatially inhomogeneous magnetic field in $x$-direction acting on the spins. Interestingly, this specific choice of the perturbation leaves the Ising stationary configurations of Sec.\ \ref{sec_Ising} unaffected: Every Ising configuration $(\vartheta_1^\text{s},\dotsc,\vartheta_N^\text{s})$ with $\vartheta_i^\text{s}\in\{0,\pi\}$ is also a stationary point of the perturbed Hamiltonian \eqref{cos_pert} for arbitrary perturbation fields $h_i$. Mathematically, this is due to the fact that the Taylor expansion of the perturbation around an Ising stationary configuration
\beq
\sum_{i = 1}^N h_i\cos\vartheta_i\Big|_{\vartheta_i=\vartheta_i^\text{s}} = \sum_{i = 1}^N \left[h_i\cos\vartheta_i^\text{s} + \mathcal{O}\left(\vartheta_i-\vartheta_i^\text{s}\right)^2\right]
\eeq
has vanishing linear contributions, thus leaving these stationary points unaffected. It is unclear to the authors whether there is any physical significance to this observation. This notwithstanding, this property can be used to check if all the singular solutions of $\mathcal{H}^{(2)}$ are indeed destroyed by the perturbation \eqref{cos_pert} also in lattices larger than $3\times3$. For simplicity in the following we will restrict ourselves to two-dimensional square lattices, but we have checked that the conclusions remain valid in three dimensions.

In order to study the effect of the perturbation \eqref{cos_pert} on singular configurations, we want to construct a sample of such configurations, spread over a range of energies similar to the nonsingular ones in Fig.\ \ref{fig_ising_xy2d}. One possible strategy to do so is to take the nonsingular sample as a starting point and transform each of the configurations into a singular one by imprinting the mask
\begin{equation}\label{e:mask}
\renewcommand{\arraystretch}{1.37}
\begin{matrix}
\downarrow & \cdot & \cdot & \cdot & \cdot & \cdot & \downarrow & \uparrow\\
\cdot & \cdot & \cdot & \cdot & \cdot & \uparrow & \downarrow & \uparrow\\
\cdot & \cdot & \cdot & \cdot & \downarrow & \uparrow & \downarrow & \cdot\\
\cdot & \cdot & \cdot & \uparrow & \downarrow & \uparrow & \cdot & \cdot\\
\cdot & \cdot & \downarrow & \uparrow & \downarrow & \cdot & \cdot & \cdot\\
\cdot & \uparrow & \downarrow & \uparrow & \cdot & \cdot & \cdot & \cdot\\
\downarrow & \uparrow & \downarrow & \cdot & \cdot & \cdot & \cdot & \cdot\\
\downarrow & \uparrow & \cdot & \cdot & \cdot & \cdot & \cdot & \uparrow\\[-4mm]
\phantom{\leftarrow} & \phantom{\leftarrow} & \phantom{\leftarrow} & \phantom{\leftarrow} & \phantom{\leftarrow} & \phantom{\leftarrow} & \phantom{\leftarrow} & \phantom{\leftarrow}
\end{matrix}
\end{equation}
(or a similar one for other lattice sizes), i.e., by rotating all spins of the configuration into the orientation indicated in \eqref{e:mask}, while leaving unchanged all sites indicated by dots. The configuration in \eqref{e:mask} is similar to the one in \eqref{e:ssp}, only that the spins on the diagonal are rotated by $\pi/2$. Such a configuration, as explained in Sec.\ \ref{sec_flat}, is also singular, and it preserves the Ising-type character of the configuration. Imprinting the mask \eqref{e:mask} causes only a subextensive change of energy, and the distribution in energy density of the stationary points will therefore be similar to the distribution of the original (nonsingular) sample. Switching on the perturbation fields $h_i$ in \eqref{cos_pert} should turn all singular solutions into regular ones, and it is this effect we want to study.

\begin{figure}
\centering
\includegraphics[scale=0.2]{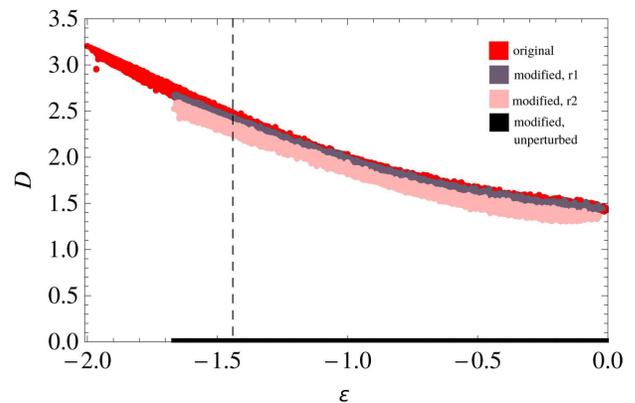}
\caption{(Color online) Rescaled Hessian determinant $D$ as a function of the energy density for the two-dimensional $XY$ model with $L=24$ and cosine perturbation terms \eqref{cos_pert}. See the legend in the plot for the color/grayscale code: ``original'' stands for the original Ising configurations as in Fig.\ \ref{fig_ising_xy2d}; ``modified'' stands for the singular Ising configurations (built from the original ones as described in the text), either with unperturbed Hamiltonian, so that $D = 0$, or with a perturbed Hamiltonian with the fields $h_i$ chosen randomly in the range $r_{1}=[-0.5,0.5]$ or $r_2=[-10^{-7},10^{-7}]$.}
\label{h_plot_r2}
\end{figure}

We performed the above analysis on $25000$ configurations for a square lattice of side length $L=24$. The fields $h_i$ were chosen randomly in the ranges $r_1=[-0.5,0.5]$ and $r_2=[-10^{-7},10^{-7}]$, to test the dependency of the reduced determinant on the strength of the fields $h_{i}$. Results are shown in Fig. \ref{h_plot_r2}.
This analysis, like the one conducted for the $3\times3$ lattice by numerical homotopy continuation, confirms that generic perturbations as in \eqref{cos_pert} transform singular solutions of $\mathcal{H}^{(2)}$ into nonsingular ones. Remarkably, the effect of the perturbations $h_i$ on the rescaled determinant $D$ is rather drastic: Already for tiny perturbations in the range $r_2=[-10^{-7},10^{-7}]$, $D$ is far away from zero and very close to the values of the original (nonsingular) Ising stationary configurations. This finding can be explained by the fact that, according to the scheme in Sec.\ \ref{sec_flat}, we constructed one-parameter families of singular solutions. Accordingly, the Hessian determinant is expected to have a single vanishing eigenvalue. Switching on a perturbation affects the zero eigenvalue by making it nonzero and of order $h$, while all other eigenvalues remain constant (nonzero) to leading order. We therefore have
\beq\label{e:Dh}
D(h) = |ch|^{1/N}\left|\prod_{k = 2}^N \lambda_k \right|^{1/N},
\eeq
where the eigenvalues $\lambda_k$ are independent of $h$ to leading order for all $k\geqslant2$. No matter how small $ch$ is, $|ch|^{1/N}$ will always be close to 1 for $N\gg1$. The remaining $(N-1)$-fold product in \eqref{e:Dh} will generically yield a value close to the `thermodynamic' one observed for generic Ising stationary configurations as shown in Fig.\ \ref{fig_ising_xy2d}, even for tiny perturbations $h$. Intuitively, we would expect a stationary point with an extensive number of vanishing eigenvalues to be more relevant for the system's thermodynamic properties, while those with a few such eigendirections should not play a major role. But this is speculation going beyond what the determinant criterion \eqref{KSS} claims and needs further examination.

In summary, we find that a generic perturbation as in \eqref{lin_pert} or \eqref{cos_pert} successfully destroys all singular stationary points. Moreover, the rescaled Hessian determinant $D$ is rather insensitive to the actual strength of the perturbation. Similar behavior is observed for the three-dimensional $XY$ model, but the results are not shown here.

\section{Conclusions}
\label{sec_conclusions}

We have explored the energy landscape of the $XY$ model with nearest-neighbor interactions on the two-dimensional square lattice and the three-dimensional cubic lattice. In particular, we have constructed certain classes of stationary points of the Hamiltonian \eqref{H_XY}. One of these classes consists of Ising stationary configurations \eqref{e:Ising}, and their number is $2^N$ for a given lattice size $N$. While analytic expressions for all these exponentially many stationary points are readily obtained, an analysis of their properties is a much harder task. We resorted to Monte Carlo techniques for generating samples of Ising stationary configurations and then numerically calculated properties like the index $\iota$ and the rescaled Hessian determinant $D$ of these points. The results, summarized in Figs.\ \ref{fig_ising_xy2d}--\ref{index_3d}, indicate that $D$ and $\iota$ are good thermodynamic observables in the sense that, with increasing lattice size $N$, the data points concentrate on a line in these plots and appear to be functions of the energy density $\varepsilon$ alone.

The original motivation for undertaking this energy landscape study was to test whether the criterion \eqref{KSS}, based on the Hessian determinant at stationary points of the Hamiltonian, reveals a signature of the phase transition of the $XY$ model in two or three dimensions. In this respect, our results are not conclusive. The data for the rescaled Hessian determinant $D$, shown in Figs.\ \ref{fig_ising_xy2d} and \ref{fig_ising_xy3d}, are clearly bounded away from zero for all values of the energy density $\varepsilon$, and therefore do not signal the presence of a phase transition according to the criterion \eqref{KSS}. As far as the validity of \eqref{KSS} is concerned, however, this finding has little to say. It rather reveals the limitations of the numerical method we have been using: The Monte Carlo technique we have been using to generate a sample of Ising stationary configurations uses importance sampling with respect to the energy, resulting in a reasonably uniform distribution of data points on the energy axis in Figs.\ \ref{fig_ising_xy2d}--\ref{index_3d}. But for a given energy density $\varepsilon$, stationary points are selected unbiased, resulting in the above mentioned behavior as `good thermodynamic observables'. This implies, however, that stationary points with vanishing (or at least small) rescaled Hessian determinant $D$ are found by this sampling technique only in case that $D=0$ is the most probable value at some energy density $\varepsilon$ (see \cite{MehtaHauensteinKastner12} for a numerical study of the nearest-neighbor $\phi^4$ model on the square lattice reaching similar conclusions). According to our data, this is not the case. 

Indeed, and rather surprisingly, we were able to show that singular stationary points, i.e., stationary points with $D=0$, do exist and are even in abundance: As proved in Sec.\ \ref{sec_flat}, even after breaking the global O(2) invariance of the $XY$ model by fixing one spin, an exponentially (in $N$) large number of singular stationary points exists, densely covering the accessible range $[-d,d]$ of energy densities in the large-$N$ limit. Moreover, these singular stationary points are non-isolated, i.e., they come in continuous families parametrized by one or several angular variables. But despite their ubiquitous presence and abundance, our Monte Carlo scheme failed to detect these points, as the value $D=0$ of their rescaled Hessian determinant is not the most probable one at any given $\varepsilon$. It must be noted that this is not a limitation of the specific Monte Carlo technique we used here: it is expected to be a generic property of unbiased numerical sampling schemes. For instance, also a search of stationary points by means of a modified Newton-Raphson method analogous to that used in Ref.\ \cite{MehtaHauensteinKastner12} did not reveal any tendency of $D$ to vanish close to the phase transition energies but did not find any singular solutions either (we have not shown these data in the paper). Hence our results suggest that from a practical point of view a purely numerical approach to the criterion \eqref{KSS} is not very useful unless a numerical sampling scheme able to efficiently detect stationary configurations with zero---or at least small---determinant is devised, which is currently lacking.

In addition to hinting at the inadequacy of commonly used numerical schemes to yield a sufficiently accurate exploration of the energy landscape of $XY$ models from the point of view of the determinant criterion, the presence of singular, non-isolated stationary points (even after explicitly breaking the global O(2) symmetry by fixing one spin) has another relevant consequence. It implies that requirements for the validity of the determinant criterion \eqref{KSS} itself, as well as of the other theoretical tools developed for the study of phase transitions based on stationary points of the energy landscape, are not met by the $XY$ Hamiltonian \eqref{H_XY}. Indeed, all these tools require that stationary points are nonsingular and isolated. This is typically assumed to be a `safe' hypothesis once global invariances of the Hamiltonian have been removed, but our results show that this is not the case.

This observation may suggest that the application of theoretical tools based on the assumption of isolated, nonsingular stationary points is hopeless in the case of $XY$ models. This is not necessarily true, because a way out consists in adding a generic perturbation to the Hamiltonian. 
We have shown in Sec.\ \ref{sec_field} that the singular stationary points can be removed by applying generic perturbations like \eqref{lin_pert} or \eqref{cos_pert}. More precisely, the removal of all the singular stationary configurations has been shown for small lattices by the homotopy continuation method. For larger lattices we have considered a sample of Ising stationary configurations, that would be singular in absence of the perturbation and that remain stationary also in presence of a perturbation of the form \eqref{cos_pert}, and we have shown that they become nonsingular when the Hamiltonian is perturbed. In previous works both the fixing of a single degree of freedom and the application of a perturbation, typically like \eqref{cos_pert} but with a homogeneous field $h$ (see e.g.\ Ref.\ \cite{jsp2003}), have been considered and were thought to be equally effective in removing singular, non-isolated solutions. Our results show that the global $O(2)$ symmetry is not the only cause of singular solutions, and the `local' strategy is therefore not sufficient for destroying them.

Remarkably, after switching on even a tiny perturbation, the rescaled determinant immediately takes on values in the vicinity of the thermodynamic average, far from the singular behavior with $D=0$. This result tells us that the study of the rescaled Hessian determinant $D$ carried out in Sec.\ \ref{sec_Ising_A} directly gives us information on the behavior of $D$ for the perturbed Hamiltonian \eqref{cos_pert} in the limit of very small external fields. Since we can now safely assume that the perturbed Hamiltonian has only isolated singular points, the results shown in Figs.\ \ref{fig_ising_xy2d} and \ref{fig_ising_xy3d} should be a faithful representation of what can be learned by standard unbiased numerical techniques as those employed in this work.

The presence of families of non-isolated stationary configurations with zero Hessian determinant (even after breaking the global O(2) symmetry) has interesting implications reaching beyond the $XY$ models investigated in the present paper. Previously, non-isolated stationary configurations had already been found in the mean-field $XY$ model, but only at a specific value of the energy density \cite{jsp2003}, and also in the globally coupled Kuramoto model with homogeneous frequencies \cite{strogatz1991stability} (in this context a continuous family of singular solutions has been termed an `incoherent manifold'). A numerical check by means of the homotopy continuation method gave similar results for a variety of other models, including the mean-field spherical $p$-spin model, particles interacting via a Lennard-Jones potential and the generalized Thomson problem, details of which will be reported elsewhere. Isolated and non-isolated singular solutions often play relevant roles also in field theories (see e.g.\ Ref.\ \cite{MehtaKastner:annphys2011} and references therein). 

\acknowledgments
R.N.\ and L.C.\ would like to warmly thank Cesare Nardini for many useful discussions. M.K.\ acknowledges support by the Incentive Funding for Rated Researchers programme of the National Research Foundation of South Africa.
D.M.\ was supported by the U.S.\ Department of Energy under Contract No.\ DE-FG02-85ER40237; would like to thank Timothy McCoy and Wenrui Hao for allowing him to use their unpublished code to verify the non-existence of the solution curves in the perturbed systems and Steven Strogatz for helpful discussions.

\appendix

\section{Singular stationary points of the three-dimensional \texorpdfstring{$XY$}{XY} model}
\label{s:A}

Similar considerations as for the two-dimensional case in Sec.\ \ref{sec_flat} motivate the following construction of singular stationary configurations in three dimensions, which for illustrational purposes is shown here for a lattice of side length $L=8$. The scheme consists of four different planar configurations 
\renewcommand{\arraystretch}{1.37}
\begin{equation}
A=\begin{matrix}
\leftarrow & \uparrow & \deemph\updownarrow & \deemph\updownarrow & \deemph\updownarrow & \deemph\updownarrow & \deemph\updownarrow & \uparrow\\
\downarrow & \deemph\updownarrow & \deemph\updownarrow & \deemph\updownarrow & \deemph\updownarrow & \deemph\updownarrow & \downarrow & \rightarrow\\
\deemph\updownarrow & \deemph\updownarrow & \deemph\updownarrow & \deemph\updownarrow & \deemph\updownarrow & \uparrow & \leftarrow & \uparrow\\
\deemph\updownarrow & \deemph\updownarrow & \deemph\updownarrow & \deemph\updownarrow & \downarrow & \rightarrow & \downarrow & \deemph\updownarrow\\
\deemph\updownarrow & \deemph\updownarrow & \deemph\updownarrow & \uparrow & \leftarrow & \uparrow & \deemph\updownarrow & \deemph\updownarrow\\
\deemph\updownarrow & \deemph\updownarrow & \downarrow & \rightarrow & \downarrow & \deemph\updownarrow & \deemph\updownarrow & \deemph\updownarrow\\
\deemph\updownarrow & \uparrow & \leftarrow & \uparrow & \deemph\updownarrow & \deemph\updownarrow & \deemph\updownarrow & \deemph\updownarrow\\
\downarrow & \rightarrow & \downarrow & \deemph\updownarrow & \deemph\updownarrow & \deemph\updownarrow & \deemph\updownarrow & \deemph\updownarrow
\end{matrix},
\end{equation}
\begin{equation}
B=\begin{matrix}
\downarrow & \deemph\updownarrow & \deemph\updownarrow & \deemph\updownarrow & \deemph\updownarrow & \deemph\updownarrow & \downarrow & \rightarrow\\
\deemph\updownarrow & \deemph\updownarrow & \deemph\updownarrow & \deemph\updownarrow & \deemph\updownarrow & \uparrow & \leftarrow & \uparrow\\
\deemph\updownarrow & \deemph\updownarrow & \deemph\updownarrow & \deemph\updownarrow & \downarrow & \rightarrow & \downarrow & \deemph\updownarrow\\
\deemph\updownarrow & \deemph\updownarrow & \deemph\updownarrow & \uparrow & \leftarrow & \uparrow & \deemph\updownarrow & \deemph\updownarrow\\
\deemph\updownarrow & \deemph\updownarrow & \downarrow & \rightarrow & \downarrow & \deemph\updownarrow & \deemph\updownarrow & \deemph\updownarrow\\
\deemph\updownarrow & \uparrow & \leftarrow & \uparrow & \deemph\updownarrow & \deemph\updownarrow & \deemph\updownarrow & \deemph\updownarrow\\
\downarrow & \rightarrow & \downarrow & \deemph\updownarrow & \deemph\updownarrow & \deemph\updownarrow & \deemph\updownarrow & \deemph\updownarrow\\
\leftarrow & \uparrow & \deemph\updownarrow & \deemph\updownarrow & \deemph\updownarrow & \deemph\updownarrow & \deemph\updownarrow & \uparrow
\end{matrix},
\end{equation}
\begin{equation}
C=\begin{matrix}
\rightarrow & \downarrow & \deemph\updownarrow & \deemph\updownarrow & \deemph\updownarrow & \deemph\updownarrow & \deemph\updownarrow & \downarrow\\
\uparrow & \deemph\updownarrow & \deemph\updownarrow & \deemph\updownarrow & \deemph\updownarrow & \deemph\updownarrow & \uparrow & \leftarrow\\
\deemph\updownarrow & \deemph\updownarrow & \deemph\updownarrow & \deemph\updownarrow & \deemph\updownarrow & \downarrow & \rightarrow & \downarrow\\
\deemph\updownarrow & \deemph\updownarrow & \deemph\updownarrow & \deemph\updownarrow & \uparrow & \leftarrow & \uparrow & \deemph\updownarrow\\
\deemph\updownarrow & \deemph\updownarrow & \deemph\updownarrow & \downarrow & \rightarrow & \downarrow & \deemph\updownarrow & \deemph\updownarrow\\
\deemph\updownarrow & \deemph\updownarrow & \uparrow & \leftarrow & \uparrow & \deemph\updownarrow & \deemph\updownarrow & \deemph\updownarrow\\
\deemph\updownarrow & \downarrow & \rightarrow & \downarrow & \deemph\updownarrow & \deemph\updownarrow & \deemph\updownarrow & \deemph\updownarrow\\
\uparrow & \leftarrow & \uparrow & \deemph\updownarrow & \deemph\updownarrow & \deemph\updownarrow & \deemph\updownarrow & \deemph\updownarrow
\end{matrix},
\end{equation}
\begin{equation}
D=\begin{matrix}
\uparrow & \deemph\updownarrow & \deemph\updownarrow & \deemph\updownarrow & \deemph\updownarrow & \deemph\updownarrow & \uparrow & \leftarrow\\
\deemph\updownarrow & \deemph\updownarrow & \deemph\updownarrow & \deemph\updownarrow & \deemph\updownarrow & \downarrow & \rightarrow & \downarrow\\
\deemph\updownarrow & \deemph\updownarrow & \deemph\updownarrow & \deemph\updownarrow & \uparrow & \leftarrow & \uparrow & \deemph\updownarrow\\
\deemph\updownarrow & \deemph\updownarrow & \deemph\updownarrow & \downarrow & \rightarrow & \downarrow & \deemph\updownarrow & \deemph\updownarrow\\
\deemph\updownarrow & \deemph\updownarrow & \uparrow & \leftarrow & \uparrow & \deemph\updownarrow & \deemph\updownarrow & \deemph\updownarrow\\
\deemph\updownarrow & \downarrow & \rightarrow & \downarrow & \deemph\updownarrow & \deemph\updownarrow & \deemph\updownarrow & \deemph\updownarrow\\
\uparrow & \leftarrow & \uparrow & \deemph\updownarrow & \deemph\updownarrow & \deemph\updownarrow & \deemph\updownarrow & \deemph\updownarrow\\
\rightarrow & \downarrow & \deemph\updownarrow & \deemph\updownarrow & \deemph\updownarrow & \deemph\updownarrow & \deemph\updownarrow & \downarrow
\end{matrix}.
\end{equation}
All these planar configurations consist of a pattern similar to the two-dimensional configuration \eqref{e:ssp}, but in $A$ and $C$ the pattern is shifted one site away from the diagonal. Moreover, $C$ is obtained from $A$ by rotating all spins by $\pi$, and the same is true for $D$ and $B$. As before, the lattice sites marked with gray $\updownarrow$-arrows can be filled with an arbitrary `Ising-type'-pattern of $\uparrow$ and $\downarrow$ arrows. Arranging these planar configurations in the sequence $ABCDABCD$ results in a stationary configuration on a cubic lattice with vanishing Hessian determinant. The scheme works in just the same way for larger lattice sizes with side lengths that are multiples of 4.

\bibliography{xy_landscape}

\begin{thebibliography}{81}%
\makeatletter
\providecommand \@ifxundefined [1]{%
 \@ifx{#1\undefined}
}%
\providecommand \@ifnum [1]{%
 \ifnum #1\expandafter \@firstoftwo
 \else \expandafter \@secondoftwo
 \fi
}%
\providecommand \@ifx [1]{%
 \ifx #1\expandafter \@firstoftwo
 \else \expandafter \@secondoftwo
 \fi
}%
\providecommand \natexlab [1]{#1}%
\providecommand \enquote  [1]{``#1''}%
\providecommand \bibnamefont  [1]{#1}%
\providecommand \bibfnamefont [1]{#1}%
\providecommand \citenamefont [1]{#1}%
\providecommand \href@noop [0]{\@secondoftwo}%
\providecommand \href [0]{\begingroup \@sanitize@url \@href}%
\providecommand \@href[1]{\@@startlink{#1}\@@href}%
\providecommand \@@href[1]{\endgroup#1\@@endlink}%
\providecommand \@sanitize@url [0]{\catcode `\\12\catcode `\$12\catcode
  `\&12\catcode `\#12\catcode `\^12\catcode `\_12\catcode `\%12\relax}%
\providecommand \@@startlink[1]{}%
\providecommand \@@endlink[0]{}%
\providecommand \url  [0]{\begingroup\@sanitize@url \@url }%
\providecommand \@url [1]{\endgroup\@href {#1}{\urlprefix }}%
\providecommand \urlprefix  [0]{URL }%
\providecommand \Eprint [0]{\href }%
\providecommand \doibase [0]{http://dx.doi.org/}%
\providecommand \selectlanguage [0]{\@gobble}%
\providecommand \bibinfo  [0]{\@secondoftwo}%
\providecommand \bibfield  [0]{\@secondoftwo}%
\providecommand \translation [1]{[#1]}%
\providecommand \BibitemOpen [0]{}%
\providecommand \bibitemStop [0]{}%
\providecommand \bibitemNoStop [0]{.\EOS\space}%
\providecommand \EOS [0]{\spacefactor3000\relax}%
\providecommand \BibitemShut  [1]{\csname bibitem#1\endcsname}%
\let\auto@bib@innerbib\@empty
\bibitem [{\citenamefont {Wales}(2004)}]{Wales:book}%
  \BibitemOpen
  \bibfield  {author} {\bibinfo {author} {\bibfnamefont {D.~J.}\ \bibnamefont
  {Wales}},\ }\href@noop {} {\emph {\bibinfo {title} {Energy Landscapes}}}\
  (\bibinfo  {publisher} {Cambridge University Press},\ \bibinfo {address}
  {Cambridge},\ \bibinfo {year} {2004})\BibitemShut {NoStop}%
\bibitem [{\citenamefont {Debenedetti}\ and\ \citenamefont
  {Stillinger}(2001)}]{DebenedettiStillinger:nature2001}%
  \BibitemOpen
  \bibfield  {author} {\bibinfo {author} {\bibfnamefont {P.~G.}\ \bibnamefont
  {Debenedetti}}\ and\ \bibinfo {author} {\bibfnamefont {F.~H.}\ \bibnamefont
  {Stillinger}},\ }\href@noop {} {\bibfield  {journal} {\bibinfo  {journal}
  {Nature}\ }\textbf {\bibinfo {volume} {410}},\ \bibinfo {pages} {259}
  (\bibinfo {year} {2001})}\BibitemShut {NoStop}%
\bibitem [{\citenamefont {Sciortino}(2005)}]{Sciortino:jstat2005}%
  \BibitemOpen
  \bibfield  {author} {\bibinfo {author} {\bibfnamefont {F.}~\bibnamefont
  {Sciortino}},\ }\href@noop {} {\bibfield  {journal} {\bibinfo  {journal} {J.
  Stat. Mech: Theory Exp.}\ }\textbf {\bibinfo {volume} {2005}},\ \bibinfo
  {pages} {P05015} (\bibinfo {year} {2005})}\BibitemShut {NoStop}%
\bibitem [{\citenamefont {Onuchic}\ \emph {et~al.}(1997)\citenamefont
  {Onuchic}, \citenamefont {Luthey-Schulten},\ and\ \citenamefont
  {Wolynes}}]{OnuchicLutheyWolynes:arpc1997}%
  \BibitemOpen
  \bibfield  {author} {\bibinfo {author} {\bibfnamefont {J.~N.}\ \bibnamefont
  {Onuchic}}, \bibinfo {author} {\bibfnamefont {Z.}~\bibnamefont
  {Luthey-Schulten}}, \ and\ \bibinfo {author} {\bibfnamefont {P.~G.}\
  \bibnamefont {Wolynes}},\ }\href@noop {} {\bibfield  {journal} {\bibinfo
  {journal} {Annu. Rev. Phys. Chem.}\ }\textbf {\bibinfo {volume} {48}},\
  \bibinfo {pages} {545} (\bibinfo {year} {1997})}\BibitemShut {NoStop}%
\bibitem [{\citenamefont {Stillinger}\ and\ \citenamefont
  {Weber}(1984)}]{StillingerWeber:science1984}%
  \BibitemOpen
  \bibfield  {author} {\bibinfo {author} {\bibfnamefont {F.~H.}\ \bibnamefont
  {Stillinger}}\ and\ \bibinfo {author} {\bibfnamefont {T.~A.}\ \bibnamefont
  {Weber}},\ }\href@noop {} {\bibfield  {journal} {\bibinfo  {journal}
  {Science}\ }\textbf {\bibinfo {volume} {225}},\ \bibinfo {pages} {983}
  (\bibinfo {year} {1984})}\BibitemShut {NoStop}%
\bibitem [{\citenamefont {Stillinger}(1995)}]{Stillinger:science1995}%
  \BibitemOpen
  \bibfield  {author} {\bibinfo {author} {\bibfnamefont {F.~H.}\ \bibnamefont
  {Stillinger}},\ }\href@noop {} {\bibfield  {journal} {\bibinfo  {journal}
  {Science}\ }\textbf {\bibinfo {volume} {267}},\ \bibinfo {pages} {1935}
  (\bibinfo {year} {1995})}\BibitemShut {NoStop}%
\bibitem [{\citenamefont {Wales}(1993)}]{wales93f}%
  \BibitemOpen
  \bibfield  {author} {\bibinfo {author} {\bibfnamefont {D.~J.}\ \bibnamefont
  {Wales}},\ }\href@noop {} {\bibfield  {journal} {\bibinfo  {journal} {Mol.
  Phys.}\ }\textbf {\bibinfo {volume} {78}},\ \bibinfo {pages} {151} (\bibinfo
  {year} {1993})}\BibitemShut {NoStop}%
\bibitem [{\citenamefont {Strodel}\ and\ \citenamefont
  {Wales}(2008)}]{StrodelWales:cpl2008}%
  \BibitemOpen
  \bibfield  {author} {\bibinfo {author} {\bibfnamefont {B.}~\bibnamefont
  {Strodel}}\ and\ \bibinfo {author} {\bibfnamefont {D.~J.}\ \bibnamefont
  {Wales}},\ }\href {\doibase 10.1016/j.cplett.2008.10.085} {\bibfield
  {journal} {\bibinfo  {journal} {Chemical Physics Letters}\ }\textbf {\bibinfo
  {volume} {466}},\ \bibinfo {pages} {105 } (\bibinfo {year}
  {2008})}\BibitemShut {NoStop}%
\bibitem [{Note1()}]{Note1}%
  \BibitemOpen
  \bibinfo {note} {The index of a stationary point $p^\protect \text {s} \in M$
  of a function $f : M \DOTSB \mapstochar \rightarrow \protect \mathbb {R}$ is
  the number of unstable directions, i.e., the number of negative eigenvalues
  of the Hessian of $f$ at $p^\protect \text {s}$.}\BibitemShut {Stop}%
\bibitem [{\citenamefont {Angelani}\ \emph {et~al.}(2000)\citenamefont
  {Angelani}, \citenamefont {{Di Leonardo}}, \citenamefont {Ruocco},
  \citenamefont {Scala},\ and\ \citenamefont
  {Sciortino}}]{AngelaniEtAl:prl2000}%
  \BibitemOpen
  \bibfield  {author} {\bibinfo {author} {\bibfnamefont {L.}~\bibnamefont
  {Angelani}}, \bibinfo {author} {\bibfnamefont {R.}~\bibnamefont {{Di
  Leonardo}}}, \bibinfo {author} {\bibfnamefont {G.}~\bibnamefont {Ruocco}},
  \bibinfo {author} {\bibfnamefont {A.}~\bibnamefont {Scala}}, \ and\ \bibinfo
  {author} {\bibfnamefont {F.}~\bibnamefont {Sciortino}},\ }\href {\doibase
  10.1103/PhysRevLett.85.5356} {\bibfield  {journal} {\bibinfo  {journal}
  {Phys. Rev. Lett.}\ }\textbf {\bibinfo {volume} {85}},\ \bibinfo {pages}
  {5356} (\bibinfo {year} {2000})}\BibitemShut {NoStop}%
\bibitem [{\citenamefont {Grigera}\ \emph {et~al.}(2002)\citenamefont
  {Grigera}, \citenamefont {Cavagna}, \citenamefont {Giardina},\ and\
  \citenamefont {Parisi}}]{GrigeraCavagnaGiardinaParisi:prl2002}%
  \BibitemOpen
  \bibfield  {author} {\bibinfo {author} {\bibfnamefont {T.~S.}\ \bibnamefont
  {Grigera}}, \bibinfo {author} {\bibfnamefont {A.}~\bibnamefont {Cavagna}},
  \bibinfo {author} {\bibfnamefont {I.}~\bibnamefont {Giardina}}, \ and\
  \bibinfo {author} {\bibfnamefont {G.}~\bibnamefont {Parisi}},\ }\href
  {\doibase 10.1103/PhysRevLett.88.055502} {\bibfield  {journal} {\bibinfo
  {journal} {Phys. Rev. Lett.}\ }\textbf {\bibinfo {volume} {88}},\ \bibinfo
  {pages} {055502} (\bibinfo {year} {2002})}\BibitemShut {NoStop}%
\bibitem [{\citenamefont {Caiani}\ \emph {et~al.}(1997)\citenamefont {Caiani},
  \citenamefont {Casetti}, \citenamefont {Clementi},\ and\ \citenamefont
  {Pettini}}]{prl1997}%
  \BibitemOpen
  \bibfield  {author} {\bibinfo {author} {\bibfnamefont {L.}~\bibnamefont
  {Caiani}}, \bibinfo {author} {\bibfnamefont {L.}~\bibnamefont {Casetti}},
  \bibinfo {author} {\bibfnamefont {C.}~\bibnamefont {Clementi}}, \ and\
  \bibinfo {author} {\bibfnamefont {M.}~\bibnamefont {Pettini}},\ }\href
  {\doibase 10.1103/PhysRevLett.79.4361} {\bibfield  {journal} {\bibinfo
  {journal} {Phys. Rev. Lett.}\ }\textbf {\bibinfo {volume} {79}},\ \bibinfo
  {pages} {4361} (\bibinfo {year} {1997})}\BibitemShut {NoStop}%
\bibitem [{\citenamefont {Casetti}\ \emph {et~al.}(1999)\citenamefont
  {Casetti}, \citenamefont {Cohen},\ and\ \citenamefont {Pettini}}]{prl1999}%
  \BibitemOpen
  \bibfield  {author} {\bibinfo {author} {\bibfnamefont {L.}~\bibnamefont
  {Casetti}}, \bibinfo {author} {\bibfnamefont {E.~G.~D.}\ \bibnamefont
  {Cohen}}, \ and\ \bibinfo {author} {\bibfnamefont {M.}~\bibnamefont
  {Pettini}},\ }\href {\doibase 10.1103/PhysRevLett.82.4160} {\bibfield
  {journal} {\bibinfo  {journal} {Phys. Rev. Lett.}\ }\textbf {\bibinfo
  {volume} {82}},\ \bibinfo {pages} {4160} (\bibinfo {year}
  {1999})}\BibitemShut {NoStop}%
\bibitem [{\citenamefont {Casetti}\ \emph {et~al.}(2000)\citenamefont
  {Casetti}, \citenamefont {Pettini},\ and\ \citenamefont
  {Cohen}}]{physrep2000}%
  \BibitemOpen
  \bibfield  {author} {\bibinfo {author} {\bibfnamefont {L.}~\bibnamefont
  {Casetti}}, \bibinfo {author} {\bibfnamefont {M.}~\bibnamefont {Pettini}}, \
  and\ \bibinfo {author} {\bibfnamefont {E.~G.~D.}\ \bibnamefont {Cohen}},\
  }\href {\doibase DOI: 10.1016/S0370-1573(00)00069-7} {\bibfield  {journal}
  {\bibinfo  {journal} {Phys. Rep.}\ }\textbf {\bibinfo {volume} {337}},\
  \bibinfo {pages} {237} (\bibinfo {year} {2000})}\BibitemShut {NoStop}%
\bibitem [{\citenamefont {Casetti}\ \emph {et~al.}(2002)\citenamefont
  {Casetti}, \citenamefont {Cohen},\ and\ \citenamefont {Pettini}}]{pre2002}%
  \BibitemOpen
  \bibfield  {author} {\bibinfo {author} {\bibfnamefont {L.}~\bibnamefont
  {Casetti}}, \bibinfo {author} {\bibfnamefont {E.~G.~D.}\ \bibnamefont
  {Cohen}}, \ and\ \bibinfo {author} {\bibfnamefont {M.}~\bibnamefont
  {Pettini}},\ }\href {\doibase 10.1103/PhysRevE.65.036112} {\bibfield
  {journal} {\bibinfo  {journal} {Phys. Rev. E}\ }\textbf {\bibinfo {volume}
  {65}},\ \bibinfo {pages} {036112} (\bibinfo {year} {2002})}\BibitemShut
  {NoStop}%
\bibitem [{\citenamefont {Casetti}\ \emph {et~al.}(2003)\citenamefont
  {Casetti}, \citenamefont {Pettini},\ and\ \citenamefont {Cohen}}]{jsp2003}%
  \BibitemOpen
  \bibfield  {author} {\bibinfo {author} {\bibfnamefont {L.}~\bibnamefont
  {Casetti}}, \bibinfo {author} {\bibfnamefont {M.}~\bibnamefont {Pettini}}, \
  and\ \bibinfo {author} {\bibfnamefont {E.~G.~D.}\ \bibnamefont {Cohen}},\
  }\href@noop {} {\bibfield  {journal} {\bibinfo  {journal} {J. Stat. Phys.}\
  }\textbf {\bibinfo {volume} {111}},\ \bibinfo {pages} {1091} (\bibinfo {year}
  {2003})}\BibitemShut {NoStop}%
\bibitem [{\citenamefont {Angelani}\ \emph {et~al.}(2003)\citenamefont
  {Angelani}, \citenamefont {Casetti}, \citenamefont {Pettini}, \citenamefont
  {Ruocco},\ and\ \citenamefont {Zamponi}}]{epl2003}%
  \BibitemOpen
  \bibfield  {author} {\bibinfo {author} {\bibfnamefont {L.}~\bibnamefont
  {Angelani}}, \bibinfo {author} {\bibfnamefont {L.}~\bibnamefont {Casetti}},
  \bibinfo {author} {\bibfnamefont {M.}~\bibnamefont {Pettini}}, \bibinfo
  {author} {\bibfnamefont {G.}~\bibnamefont {Ruocco}}, \ and\ \bibinfo {author}
  {\bibfnamefont {F.}~\bibnamefont {Zamponi}},\ }\href@noop {} {\bibfield
  {journal} {\bibinfo  {journal} {Europhys. Lett.}\ }\textbf {\bibinfo {volume}
  {62}},\ \bibinfo {pages} {775} (\bibinfo {year} {2003})}\BibitemShut
  {NoStop}%
\bibitem [{\citenamefont {Garanin}\ \emph {et~al.}(2004)\citenamefont
  {Garanin}, \citenamefont {Schilling},\ and\ \citenamefont
  {Scala}}]{GaraninSchillingScala:pre2004}%
  \BibitemOpen
  \bibfield  {author} {\bibinfo {author} {\bibfnamefont {D.~A.}\ \bibnamefont
  {Garanin}}, \bibinfo {author} {\bibfnamefont {R.}~\bibnamefont {Schilling}},
  \ and\ \bibinfo {author} {\bibfnamefont {A.}~\bibnamefont {Scala}},\ }\href
  {\doibase 10.1103/PhysRevE.70.036125} {\bibfield  {journal} {\bibinfo
  {journal} {Phys. Rev. E}\ }\textbf {\bibinfo {volume} {70}},\ \bibinfo
  {pages} {036125} (\bibinfo {year} {2004})}\BibitemShut {NoStop}%
\bibitem [{\citenamefont {Grinza}\ and\ \citenamefont
  {Mossa}(2004)}]{GrinzaMossa:prl2004}%
  \BibitemOpen
  \bibfield  {author} {\bibinfo {author} {\bibfnamefont {P.}~\bibnamefont
  {Grinza}}\ and\ \bibinfo {author} {\bibfnamefont {A.}~\bibnamefont {Mossa}},\
  }\href {\doibase 10.1103/PhysRevLett.92.158102} {\bibfield  {journal}
  {\bibinfo  {journal} {Phys. Rev. Lett.}\ }\textbf {\bibinfo {volume} {92}},\
  \bibinfo {pages} {158102} (\bibinfo {year} {2004})}\BibitemShut {NoStop}%
\bibitem [{\citenamefont {{Ribeiro Teixeira}}\ and\ \citenamefont
  {Stariolo}(2004)}]{RibeiroStariolo:pre2004}%
  \BibitemOpen
  \bibfield  {author} {\bibinfo {author} {\bibfnamefont {A.~C.}\ \bibnamefont
  {{Ribeiro Teixeira}}}\ and\ \bibinfo {author} {\bibfnamefont {D.~A.}\
  \bibnamefont {Stariolo}},\ }\href {\doibase 10.1103/PhysRevE.70.016113}
  {\bibfield  {journal} {\bibinfo  {journal} {Phys. Rev. E}\ }\textbf {\bibinfo
  {volume} {70}},\ \bibinfo {pages} {016113} (\bibinfo {year}
  {2004})}\BibitemShut {NoStop}%
\bibitem [{\citenamefont {Andronico}\ \emph {et~al.}(2004)\citenamefont
  {Andronico}, \citenamefont {Angelani}, \citenamefont {Ruocco},\ and\
  \citenamefont {Zamponi}}]{AndronicoEtAl:pre2004}%
  \BibitemOpen
  \bibfield  {author} {\bibinfo {author} {\bibfnamefont {A.}~\bibnamefont
  {Andronico}}, \bibinfo {author} {\bibfnamefont {L.}~\bibnamefont {Angelani}},
  \bibinfo {author} {\bibfnamefont {G.}~\bibnamefont {Ruocco}}, \ and\ \bibinfo
  {author} {\bibfnamefont {F.}~\bibnamefont {Zamponi}},\ }\href {\doibase
  10.1103/PhysRevE.70.041101} {\bibfield  {journal} {\bibinfo  {journal} {Phys.
  Rev. E}\ }\textbf {\bibinfo {volume} {70}},\ \bibinfo {pages} {041101}
  (\bibinfo {year} {2004})}\BibitemShut {NoStop}%
\bibitem [{\citenamefont {Kastner}(2004)}]{Kastner:prl2004}%
  \BibitemOpen
  \bibfield  {author} {\bibinfo {author} {\bibfnamefont {M.}~\bibnamefont
  {Kastner}},\ }\href {\doibase 10.1103/PhysRevLett.93.150601} {\bibfield
  {journal} {\bibinfo  {journal} {Phys. Rev. Lett.}\ }\textbf {\bibinfo
  {volume} {93}},\ \bibinfo {pages} {150601} (\bibinfo {year}
  {2004})}\BibitemShut {NoStop}%
\bibitem [{\citenamefont {Angelani}\ \emph
  {et~al.}(2005{\natexlab{a}})\citenamefont {Angelani}, \citenamefont
  {Casetti}, \citenamefont {Pettini}, \citenamefont {Ruocco},\ and\
  \citenamefont {Zamponi}}]{pre2005}%
  \BibitemOpen
  \bibfield  {author} {\bibinfo {author} {\bibfnamefont {L.}~\bibnamefont
  {Angelani}}, \bibinfo {author} {\bibfnamefont {L.}~\bibnamefont {Casetti}},
  \bibinfo {author} {\bibfnamefont {M.}~\bibnamefont {Pettini}}, \bibinfo
  {author} {\bibfnamefont {G.}~\bibnamefont {Ruocco}}, \ and\ \bibinfo {author}
  {\bibfnamefont {F.}~\bibnamefont {Zamponi}},\ }\href {\doibase
  10.1103/PhysRevE.71.036152} {\bibfield  {journal} {\bibinfo  {journal} {Phys.
  Rev. E}\ }\textbf {\bibinfo {volume} {71}},\ \bibinfo {pages} {036152}
  (\bibinfo {year} {2005}{\natexlab{a}})}\BibitemShut {NoStop}%
\bibitem [{\citenamefont {Angelani}\ \emph
  {et~al.}(2005{\natexlab{b}})\citenamefont {Angelani}, \citenamefont
  {Ruocco},\ and\ \citenamefont {Zamponi}}]{AngelaniRuoccoZamponi:pre2005}%
  \BibitemOpen
  \bibfield  {author} {\bibinfo {author} {\bibfnamefont {L.}~\bibnamefont
  {Angelani}}, \bibinfo {author} {\bibfnamefont {G.}~\bibnamefont {Ruocco}}, \
  and\ \bibinfo {author} {\bibfnamefont {F.}~\bibnamefont {Zamponi}},\ }\href
  {\doibase 10.1103/PhysRevE.72.016122} {\bibfield  {journal} {\bibinfo
  {journal} {Phys. Rev. E}\ }\textbf {\bibinfo {volume} {72}},\ \bibinfo
  {pages} {016122} (\bibinfo {year} {2005}{\natexlab{b}})}\BibitemShut
  {NoStop}%
\bibitem [{\citenamefont {Risau-Gusman}\ \emph {et~al.}(2005)\citenamefont
  {Risau-Gusman}, \citenamefont {Ribeiro-Teixeira},\ and\ \citenamefont
  {Stariolo}}]{Risau-GusmanEtAl:prl2005}%
  \BibitemOpen
  \bibfield  {author} {\bibinfo {author} {\bibfnamefont {S.}~\bibnamefont
  {Risau-Gusman}}, \bibinfo {author} {\bibfnamefont {A.~C.}\ \bibnamefont
  {Ribeiro-Teixeira}}, \ and\ \bibinfo {author} {\bibfnamefont {D.~A.}\
  \bibnamefont {Stariolo}},\ }\href {\doibase 10.1103/PhysRevLett.95.145702}
  {\bibfield  {journal} {\bibinfo  {journal} {Phys. Rev. Lett.}\ }\textbf
  {\bibinfo {volume} {95}},\ \bibinfo {pages} {145702} (\bibinfo {year}
  {2005})}\BibitemShut {NoStop}%
\bibitem [{\citenamefont {Hahn}\ and\ \citenamefont
  {Kastner}(2005)}]{HahnKastner:pre2005}%
  \BibitemOpen
  \bibfield  {author} {\bibinfo {author} {\bibfnamefont {I.}~\bibnamefont
  {Hahn}}\ and\ \bibinfo {author} {\bibfnamefont {M.}~\bibnamefont {Kastner}},\
  }\href@noop {} {\bibfield  {journal} {\bibinfo  {journal} {Phys. Rev. E}\
  }\textbf {\bibinfo {volume} {72}},\ \bibinfo {pages} {056134} (\bibinfo
  {year} {2005})}\BibitemShut {NoStop}%
\bibitem [{\citenamefont {Baroni}\ and\ \citenamefont
  {Casetti}(2006)}]{jpa2006}%
  \BibitemOpen
  \bibfield  {author} {\bibinfo {author} {\bibfnamefont {F.}~\bibnamefont
  {Baroni}}\ and\ \bibinfo {author} {\bibfnamefont {L.}~\bibnamefont
  {Casetti}},\ }\href@noop {} {\bibfield  {journal} {\bibinfo  {journal} {J.
  Phys. A: Math. Gen.}\ }\textbf {\bibinfo {volume} {39}},\ \bibinfo {pages}
  {529} (\bibinfo {year} {2006})}\BibitemShut {NoStop}%
\bibitem [{\citenamefont {Risau-Gusman}\ \emph {et~al.}(2006)\citenamefont
  {Risau-Gusman}, \citenamefont {Ribeiro-Teixeira},\ and\ \citenamefont
  {Stariolo}}]{Risau-GusmanEtAl:jsp2006}%
  \BibitemOpen
  \bibfield  {author} {\bibinfo {author} {\bibfnamefont {S.}~\bibnamefont
  {Risau-Gusman}}, \bibinfo {author} {\bibfnamefont {A.~C.}\ \bibnamefont
  {Ribeiro-Teixeira}}, \ and\ \bibinfo {author} {\bibfnamefont
  {D.}~\bibnamefont {Stariolo}},\ }\href@noop {} {\bibfield  {journal}
  {\bibinfo  {journal} {J. Stat. Phys.}\ }\textbf {\bibinfo {volume} {124}},\
  \bibinfo {pages} {1231} (\bibinfo {year} {2006})}\BibitemShut {NoStop}%
\bibitem [{\citenamefont {Angelani}\ and\ \citenamefont
  {Ruocco}(2007)}]{AngelaniRuocco:pre2007}%
  \BibitemOpen
  \bibfield  {author} {\bibinfo {author} {\bibfnamefont {L.}~\bibnamefont
  {Angelani}}\ and\ \bibinfo {author} {\bibfnamefont {G.}~\bibnamefont
  {Ruocco}},\ }\href {\doibase 10.1103/PhysRevE.76.051119} {\bibfield
  {journal} {\bibinfo  {journal} {Phys. Rev. E}\ }\textbf {\bibinfo {volume}
  {76}},\ \bibinfo {pages} {051119} (\bibinfo {year} {2007})}\BibitemShut
  {NoStop}%
\bibitem [{\citenamefont {Angelani}\ and\ \citenamefont
  {Ruocco}(2008)}]{AngelaniRuocco:pre2008}%
  \BibitemOpen
  \bibfield  {author} {\bibinfo {author} {\bibfnamefont {L.}~\bibnamefont
  {Angelani}}\ and\ \bibinfo {author} {\bibfnamefont {G.}~\bibnamefont
  {Ruocco}},\ }\href {\doibase 10.1103/PhysRevE.77.052101} {\bibfield
  {journal} {\bibinfo  {journal} {Phys. Rev. E}\ }\textbf {\bibinfo {volume}
  {77}},\ \bibinfo {pages} {052101} (\bibinfo {year} {2008})}\BibitemShut
  {NoStop}%
\bibitem [{\citenamefont {Nardini}\ and\ \citenamefont
  {Casetti}(2009)}]{prerap2009}%
  \BibitemOpen
  \bibfield  {author} {\bibinfo {author} {\bibfnamefont {C.}~\bibnamefont
  {Nardini}}\ and\ \bibinfo {author} {\bibfnamefont {L.}~\bibnamefont
  {Casetti}},\ }\href {\doibase 10.1103/PhysRevE.80.060103} {\bibfield
  {journal} {\bibinfo  {journal} {Phys. Rev. E}\ }\textbf {\bibinfo {volume}
  {80}},\ \bibinfo {pages} {060103} (\bibinfo {year} {2009})}\BibitemShut
  {NoStop}%
\bibitem [{\citenamefont {Santos}\ and\ \citenamefont
  {Coutinho-Filho}(2009)}]{SantosCoutinho-Filho:pre2009}%
  \BibitemOpen
  \bibfield  {author} {\bibinfo {author} {\bibfnamefont {F.~A.~N.}\
  \bibnamefont {Santos}}\ and\ \bibinfo {author} {\bibfnamefont {M.~D.}\
  \bibnamefont {Coutinho-Filho}},\ }\href {\doibase 10.1103/PhysRevE.80.031123}
  {\bibfield  {journal} {\bibinfo  {journal} {Phys. Rev. E}\ }\textbf {\bibinfo
  {volume} {80}},\ \bibinfo {pages} {031123} (\bibinfo {year}
  {2009})}\BibitemShut {NoStop}%
\bibitem [{\citenamefont {Kastner}(2011)}]{Kastner:pre2011}%
  \BibitemOpen
  \bibfield  {author} {\bibinfo {author} {\bibfnamefont {M.}~\bibnamefont
  {Kastner}},\ }\href {\doibase 10.1103/PhysRevE.83.031114} {\bibfield
  {journal} {\bibinfo  {journal} {Phys. Rev. E}\ }\textbf {\bibinfo {volume}
  {83}},\ \bibinfo {pages} {031114} (\bibinfo {year} {2011})}\BibitemShut
  {NoStop}%
\bibitem [{\citenamefont {Mehta}\ and\ \citenamefont
  {Kastner}(2011)}]{MehtaKastner:annphys2011}%
  \BibitemOpen
  \bibfield  {author} {\bibinfo {author} {\bibfnamefont {D.}~\bibnamefont
  {Mehta}}\ and\ \bibinfo {author} {\bibfnamefont {M.}~\bibnamefont
  {Kastner}},\ }\href {\doibase DOI: 10.1016/j.aop.2010.12.016} {\bibfield
  {journal} {\bibinfo  {journal} {Ann. Phys. (NY)}\ }\textbf {\bibinfo {volume}
  {326}},\ \bibinfo {pages} {1425} (\bibinfo {year} {2011})}\BibitemShut
  {NoStop}%
\bibitem [{\citenamefont {Mehta}(2011{\natexlab{a}})}]{Mehta:2011xs}%
  \BibitemOpen
  \bibfield  {author} {\bibinfo {author} {\bibfnamefont {D.}~\bibnamefont
  {Mehta}},\ }\href {\doibase 10.1103/PhysRevE.84.025702} {\bibfield  {journal}
  {\bibinfo  {journal} {Phys. Rev. E}\ }\textbf {\bibinfo {volume} {84}},\
  \bibinfo {pages} {025702} (\bibinfo {year} {2011}{\natexlab{a}})}\BibitemShut
  {NoStop}%
\bibitem [{\citenamefont {Baroni}(2011)}]{Baroni:jstat2011}%
  \BibitemOpen
  \bibfield  {author} {\bibinfo {author} {\bibfnamefont {F.}~\bibnamefont
  {Baroni}},\ }\href@noop {} {\bibfield  {journal} {\bibinfo  {journal} {J.
  Stat. Mech: Theory Exp.}\ }\textbf {\bibinfo {volume} {2011}},\ \bibinfo
  {pages} {P08010} (\bibinfo {year} {2011})}\BibitemShut {NoStop}%
\bibitem [{\citenamefont {Farber}\ and\ \citenamefont
  {Fromm}(2011)}]{FarberFromm:jams2011}%
  \BibitemOpen
  \bibfield  {author} {\bibinfo {author} {\bibfnamefont {M.}~\bibnamefont
  {Farber}}\ and\ \bibinfo {author} {\bibfnamefont {V.}~\bibnamefont {Fromm}},\
  }\href {\doibase 10.1017/S144678871100125X} {\bibfield  {journal} {\bibinfo
  {journal} {J. Aust. Math. Soc.}\ }\textbf {\bibinfo {volume} {90}},\ \bibinfo
  {pages} {183} (\bibinfo {year} {2011})}\BibitemShut {NoStop}%
\bibitem [{\citenamefont {Carlsson}\ \emph {et~al.}(2012)\citenamefont
  {Carlsson}, \citenamefont {Gorham}, \citenamefont {Kahle},\ and\
  \citenamefont {Mason}}]{CarlssonEtAl:pre2012}%
  \BibitemOpen
  \bibfield  {author} {\bibinfo {author} {\bibfnamefont {G.}~\bibnamefont
  {Carlsson}}, \bibinfo {author} {\bibfnamefont {J.}~\bibnamefont {Gorham}},
  \bibinfo {author} {\bibfnamefont {M.}~\bibnamefont {Kahle}}, \ and\ \bibinfo
  {author} {\bibfnamefont {J.}~\bibnamefont {Mason}},\ }\href {\doibase
  10.1103/PhysRevE.85.011303} {\bibfield  {journal} {\bibinfo  {journal} {Phys.
  Rev. E}\ }\textbf {\bibinfo {volume} {85}},\ \bibinfo {pages} {011303}
  (\bibinfo {year} {2012})}\BibitemShut {NoStop}%
\bibitem [{\citenamefont {Kastner}(2008)}]{Kastner:rmp2008}%
  \BibitemOpen
  \bibfield  {author} {\bibinfo {author} {\bibfnamefont {M.}~\bibnamefont
  {Kastner}},\ }\href {\doibase 10.1103/RevModPhys.80.167} {\bibfield
  {journal} {\bibinfo  {journal} {Rev. Mod. Phys.}\ }\textbf {\bibinfo {volume}
  {80}},\ \bibinfo {pages} {167} (\bibinfo {year} {2008})}\BibitemShut
  {NoStop}%
\bibitem [{\citenamefont {Pettini}(2007)}]{Pettini:book}%
  \BibitemOpen
  \bibfield  {author} {\bibinfo {author} {\bibfnamefont {M.}~\bibnamefont
  {Pettini}},\ }\href@noop {} {\emph {\bibinfo {title} {Geometry and Topology
  in Hamiltonian Dynamics and Statistical Mechanics}}}\ (\bibinfo  {publisher}
  {Springer},\ \bibinfo {address} {New York},\ \bibinfo {year}
  {2007})\BibitemShut {NoStop}%
\bibitem [{\citenamefont {Kastner}(2006)}]{Kastner06}%
  \BibitemOpen
  \bibfield  {author} {\bibinfo {author} {\bibfnamefont {M.}~\bibnamefont
  {Kastner}},\ }\href@noop {} {\bibfield  {journal} {\bibinfo  {journal}
  {Physica A}\ }\textbf {\bibinfo {volume} {359}},\ \bibinfo {pages} {447}
  (\bibinfo {year} {2006})}\BibitemShut {NoStop}%
\bibitem [{Note2()}]{Note2}%
  \BibitemOpen
  \bibinfo {note} {Throughout the paper we set Boltzmann's constant $k_B$ to
  unity.}\BibitemShut {Stop}%
\bibitem [{\citenamefont {Federer}(1969)}]{Federer:book}%
  \BibitemOpen
  \bibfield  {author} {\bibinfo {author} {\bibfnamefont {H.}~\bibnamefont
  {Federer}},\ }\href@noop {} {\emph {\bibinfo {title} {Geometric Measure
  Theory}}}\ (\bibinfo  {publisher} {Springer},\ \bibinfo {address} {New
  York},\ \bibinfo {year} {1969})\BibitemShut {NoStop}%
\bibitem [{\citenamefont {Kastner}\ \emph {et~al.}(2008)\citenamefont
  {Kastner}, \citenamefont {Schnetz},\ and\ \citenamefont
  {Schreiber}}]{KSS:jstat2008}%
  \BibitemOpen
  \bibfield  {author} {\bibinfo {author} {\bibfnamefont {M.}~\bibnamefont
  {Kastner}}, \bibinfo {author} {\bibfnamefont {O.}~\bibnamefont {Schnetz}}, \
  and\ \bibinfo {author} {\bibfnamefont {S.}~\bibnamefont {Schreiber}},\
  }\href@noop {} {\bibfield  {journal} {\bibinfo  {journal} {J. Stat. Mech:
  Theory Exp.}\ }\textbf {\bibinfo {volume} {2008}},\ \bibinfo {pages} {P04025}
  (\bibinfo {year} {2008})}\BibitemShut {NoStop}%
\bibitem [{Note3()}]{Note3}%
  \BibitemOpen
  \bibinfo {note} {Such a behavior differs from the canonical ensemble where
  the canonical free energy or other thermodynamic functions may develop
  nonanalyticities only in the thermodynamic limit $N\to \infty $ \cite
  {Griffiths:inDombGreen}.}\BibitemShut {Stop}%
\bibitem [{\citenamefont {Casetti}\ \emph {et~al.}(2009)\citenamefont
  {Casetti}, \citenamefont {Kastner},\ and\ \citenamefont
  {Nerattini}}]{jstat2009}%
  \BibitemOpen
  \bibfield  {author} {\bibinfo {author} {\bibfnamefont {L.}~\bibnamefont
  {Casetti}}, \bibinfo {author} {\bibfnamefont {M.}~\bibnamefont {Kastner}}, \
  and\ \bibinfo {author} {\bibfnamefont {R.}~\bibnamefont {Nerattini}},\
  }\href@noop {} {\bibfield  {journal} {\bibinfo  {journal} {J. Stat. Mech:
  Theory Exp.}\ }\textbf {\bibinfo {volume} {2009}},\ \bibinfo {pages} {P07036}
  (\bibinfo {year} {2009})}\BibitemShut {NoStop}%
\bibitem [{Note4()}]{Note4}%
  \BibitemOpen
  \bibinfo {note} {This challenge is only apparent. In general, the order of a
  nonanalyticity in the infinite system is not necessarily equal to the
  large-$N$ limit of the order of finite system nonanalyticities, as this
  identification would require two limiting procedures to commute.}\BibitemShut
  {Stop}%
\bibitem [{\citenamefont {Baroni}(2002)}]{Fabrizio:thesis}%
  \BibitemOpen
  \bibfield  {author} {\bibinfo {author} {\bibfnamefont {F.}~\bibnamefont
  {Baroni}},\ }\emph {\bibinfo {title} {Transizioni di fase e topologia dello
  spazio delle configurazioni di modelli di campo medio}},\ \href@noop {}
  {Master's thesis},\ \bibinfo  {school} {Universit{\`a} di Firenze} (\bibinfo
  {year} {2002})\BibitemShut {NoStop}%
\bibitem [{\citenamefont {Kastner}\ and\ \citenamefont
  {Mehta}(2011)}]{KastnerMehta:prl2011}%
  \BibitemOpen
  \bibfield  {author} {\bibinfo {author} {\bibfnamefont {M.}~\bibnamefont
  {Kastner}}\ and\ \bibinfo {author} {\bibfnamefont {D.}~\bibnamefont
  {Mehta}},\ }\href {\doibase 10.1103/PhysRevLett.107.160602} {\bibfield
  {journal} {\bibinfo  {journal} {Phys. Rev. Lett.}\ }\textbf {\bibinfo
  {volume} {107}},\ \bibinfo {pages} {160602} (\bibinfo {year}
  {2011})}\BibitemShut {NoStop}%
\bibitem [{Note5()}]{Note5}%
  \BibitemOpen
  \bibinfo {note} {A theorem according to which topology changes, and hence
  stationary points of the potential energy, would be a necessary condition for
  phase transition in systems with short-ranged and confining interactions has
  been announced in \cite {FranzosiPettini:prl} and its proof has been
  presented in \cite {FranzosiPettini:npb}. However, a counterexample has been
  found and discussed in \cite {KastnerMehta:prl2011}.}\BibitemShut {Stop}%
\bibitem [{\citenamefont {Casetti}\ and\ \citenamefont
  {Kastner}(2006)}]{prl2006}%
  \BibitemOpen
  \bibfield  {author} {\bibinfo {author} {\bibfnamefont {L.}~\bibnamefont
  {Casetti}}\ and\ \bibinfo {author} {\bibfnamefont {M.}~\bibnamefont
  {Kastner}},\ }\href {\doibase 10.1103/PhysRevLett.97.100602} {\bibfield
  {journal} {\bibinfo  {journal} {Phys. Rev. Lett.}\ }\textbf {\bibinfo
  {volume} {97}},\ \bibinfo {pages} {100602} (\bibinfo {year}
  {2006})}\BibitemShut {NoStop}%
\bibitem [{\citenamefont {Kastner}\ \emph {et~al.}(2007)\citenamefont
  {Kastner}, \citenamefont {Schreiber},\ and\ \citenamefont
  {Schnetz}}]{KSS:prl2007}%
  \BibitemOpen
  \bibfield  {author} {\bibinfo {author} {\bibfnamefont {M.}~\bibnamefont
  {Kastner}}, \bibinfo {author} {\bibfnamefont {S.}~\bibnamefont {Schreiber}},
  \ and\ \bibinfo {author} {\bibfnamefont {O.}~\bibnamefont {Schnetz}},\ }\href
  {\doibase 10.1103/PhysRevLett.99.050601} {\bibfield  {journal} {\bibinfo
  {journal} {Phys. Rev. Lett.}\ }\textbf {\bibinfo {volume} {99}},\ \bibinfo
  {pages} {050601} (\bibinfo {year} {2007})}\BibitemShut {NoStop}%
\bibitem [{\citenamefont {Kastner}\ and\ \citenamefont
  {Schnetz}(2008)}]{KastnerSchnetz:prl2008}%
  \BibitemOpen
  \bibfield  {author} {\bibinfo {author} {\bibfnamefont {M.}~\bibnamefont
  {Kastner}}\ and\ \bibinfo {author} {\bibfnamefont {O.}~\bibnamefont
  {Schnetz}},\ }\href {\doibase 10.1103/PhysRevLett.100.160601} {\bibfield
  {journal} {\bibinfo  {journal} {Phys. Rev. Lett.}\ }\textbf {\bibinfo
  {volume} {100}},\ \bibinfo {pages} {160601} (\bibinfo {year}
  {2008})}\BibitemShut {NoStop}%
\bibitem [{\citenamefont {Mehta}(2009)}]{Mehta:2009}%
  \BibitemOpen
  \bibfield  {author} {\bibinfo {author} {\bibfnamefont {D.}~\bibnamefont
  {Mehta}},\ }\emph {\bibinfo {title} {Lattice vs. Continuum: Landau Gauge
  Fixing and 't Hooft-Polyakov Monopoles}},\ \href@noop {} {Ph.D. thesis},\
  \bibinfo  {school} {The University of Adelaide} (\bibinfo {year} {2009}),\
  \bibinfo {note} {{A}ustralasian Digital Theses Program}\BibitemShut {NoStop}%
\bibitem [{\citenamefont {Mehta}\ \emph {et~al.}(2009)\citenamefont {Mehta},
  \citenamefont {Sternbeck}, \citenamefont {von Smekal},\ and\ \citenamefont
  {Williams}}]{Mehta:2009zv}%
  \BibitemOpen
  \bibfield  {author} {\bibinfo {author} {\bibfnamefont {D.}~\bibnamefont
  {Mehta}}, \bibinfo {author} {\bibfnamefont {A.}~\bibnamefont {Sternbeck}},
  \bibinfo {author} {\bibfnamefont {L.}~\bibnamefont {von Smekal}}, \ and\
  \bibinfo {author} {\bibfnamefont {A.~G.}\ \bibnamefont {Williams}},\
  }\href@noop {} {\bibfield  {journal} {\bibinfo  {journal} {PoS}\ }\textbf
  {\bibinfo {volume} {QCD-TNT09}},\ \bibinfo {pages} {25} (\bibinfo {year}
  {2009})}\BibitemShut {NoStop}%
\bibitem [{\citenamefont {Hughes}\ \emph {et~al.}()\citenamefont {Hughes},
  \citenamefont {Mehta},\ and\ \citenamefont {Skullerud}}]{Hughes:2012hg}%
  \BibitemOpen
  \bibfield  {author} {\bibinfo {author} {\bibfnamefont {C.}~\bibnamefont
  {Hughes}}, \bibinfo {author} {\bibfnamefont {D.}~\bibnamefont {Mehta}}, \
  and\ \bibinfo {author} {\bibfnamefont {J.-I.}\ \bibnamefont {Skullerud}},\
  }\href@noop {} {\enquote {\bibinfo {title} {{Enumerating Gribov copies on the
  lattice}},}\ }\Eprint {http://arxiv.org/abs/1203.4847} {arXiv:1203.4847}
  \BibitemShut {NoStop}%
\bibitem [{\citenamefont {Wales}\ and\ \citenamefont
  {Doye}(2003)}]{WalesDoye:jcp2003}%
  \BibitemOpen
  \bibfield  {author} {\bibinfo {author} {\bibfnamefont {D.~J.}\ \bibnamefont
  {Wales}}\ and\ \bibinfo {author} {\bibfnamefont {J.~P.~K.}\ \bibnamefont
  {Doye}},\ }\href@noop {} {\bibfield  {journal} {\bibinfo  {journal} {J. Chem.
  Phys.}\ }\textbf {\bibinfo {volume} {119}},\ \bibinfo {pages} {12409}
  (\bibinfo {year} {2003})}\BibitemShut {NoStop}%
\bibitem [{\citenamefont {Schilling}(2006)}]{Schilling:physicad2006}%
  \BibitemOpen
  \bibfield  {author} {\bibinfo {author} {\bibfnamefont {R.}~\bibnamefont
  {Schilling}},\ }\href {\doibase DOI: 10.1016/j.physd.2005.12.013} {\bibfield
  {journal} {\bibinfo  {journal} {Physica D}\ }\textbf {\bibinfo {volume}
  {216}},\ \bibinfo {pages} {157} (\bibinfo {year} {2006})}\BibitemShut
  {NoStop}%
\bibitem [{\citenamefont {Casetti}\ \emph {et~al.}(2011)\citenamefont
  {Casetti}, \citenamefont {Nardini},\ and\ \citenamefont
  {Nerattini}}]{prl2011}%
  \BibitemOpen
  \bibfield  {author} {\bibinfo {author} {\bibfnamefont {L.}~\bibnamefont
  {Casetti}}, \bibinfo {author} {\bibfnamefont {C.}~\bibnamefont {Nardini}}, \
  and\ \bibinfo {author} {\bibfnamefont {R.}~\bibnamefont {Nerattini}},\ }\href
  {\doibase 10.1103/PhysRevLett.106.057208} {\bibfield  {journal} {\bibinfo
  {journal} {Phys. Rev. Lett.}\ }\textbf {\bibinfo {volume} {106}},\ \bibinfo
  {pages} {057208} (\bibinfo {year} {2011})}\BibitemShut {NoStop}%
\bibitem [{\citenamefont {Nardini}\ \emph {et~al.}(2012)\citenamefont
  {Nardini}, \citenamefont {Nerattini},\ and\ \citenamefont
  {Casetti}}]{jstat2012}%
  \BibitemOpen
  \bibfield  {author} {\bibinfo {author} {\bibfnamefont {C.}~\bibnamefont
  {Nardini}}, \bibinfo {author} {\bibfnamefont {R.}~\bibnamefont {Nerattini}},
  \ and\ \bibinfo {author} {\bibfnamefont {L.}~\bibnamefont {Casetti}},\ }\href
  {\doibase 10.1088/1742-5468/2012/02/P02007} {\bibfield  {journal} {\bibinfo
  {journal} {J. Stat. Mech: Theory Exp.}\ }\textbf {\bibinfo {volume} {2012}},\
  \bibinfo {pages} {P02007} (\bibinfo {year} {2012})}\BibitemShut {NoStop}%
\bibitem [{\citenamefont
  {Bere\v{z}inskij}(1971)}]{berezinskij:sovphysjetp1971}%
  \BibitemOpen
  \bibfield  {author} {\bibinfo {author} {\bibfnamefont {V.~L.}\ \bibnamefont
  {Bere\v{z}inskij}},\ }\href@noop {} {\bibfield  {journal} {\bibinfo
  {journal} {Sov. Phys. JETP}\ }\textbf {\bibinfo {volume} {32}},\ \bibinfo
  {pages} {493} (\bibinfo {year} {1971})}\BibitemShut {NoStop}%
\bibitem [{\citenamefont {Kosterlitz}\ and\ \citenamefont
  {Thouless}(1973)}]{KosterlitzThouless:jphysc1973}%
  \BibitemOpen
  \bibfield  {author} {\bibinfo {author} {\bibfnamefont {J.~M.}\ \bibnamefont
  {Kosterlitz}}\ and\ \bibinfo {author} {\bibfnamefont {D.~J.}\ \bibnamefont
  {Thouless}},\ }\href@noop {} {\bibfield  {journal} {\bibinfo  {journal} {J.
  Phys. C: Solid State Phys.}\ }\textbf {\bibinfo {volume} {6}},\ \bibinfo
  {pages} {1181} (\bibinfo {year} {1973})}\BibitemShut {NoStop}%
\bibitem [{\citenamefont {Goldenfeld}(1992)}]{Goldenfeld:book}%
  \BibitemOpen
  \bibfield  {author} {\bibinfo {author} {\bibfnamefont {N.}~\bibnamefont
  {Goldenfeld}},\ }\href@noop {} {\emph {\bibinfo {title} {Lectures on Phase
  Transitions and the Renormalisation Group}}}\ (\bibinfo  {publisher} {Perseus
  Publishing},\ \bibinfo {address} {Cambridge},\ \bibinfo {year}
  {1992})\BibitemShut {NoStop}%
\bibitem [{\citenamefont {Gupta}\ and\ \citenamefont
  {Baillie}(1992)}]{GuptaBaillie:prb1992}%
  \BibitemOpen
  \bibfield  {author} {\bibinfo {author} {\bibfnamefont {R.}~\bibnamefont
  {Gupta}}\ and\ \bibinfo {author} {\bibfnamefont {C.~F.}\ \bibnamefont
  {Baillie}},\ }\href {\doibase 10.1103/PhysRevB.45.2883} {\bibfield  {journal}
  {\bibinfo  {journal} {Phys. Rev. B}\ }\textbf {\bibinfo {volume} {45}},\
  \bibinfo {pages} {2883} (\bibinfo {year} {1992})}\BibitemShut {NoStop}%
\bibitem [{\citenamefont {Hasenbusch}(2005)}]{Hasenbusch:jphysa2005}%
  \BibitemOpen
  \bibfield  {author} {\bibinfo {author} {\bibfnamefont {M.}~\bibnamefont
  {Hasenbusch}},\ }\href@noop {} {\bibfield  {journal} {\bibinfo  {journal} {J.
  Phys. A: Math. Gen.}\ }\textbf {\bibinfo {volume} {38}},\ \bibinfo {pages}
  {5869} (\bibinfo {year} {2005})}\BibitemShut {NoStop}%
\bibitem [{\citenamefont {Gottlob}\ and\ \citenamefont
  {Hasenbusch}(1993)}]{GottlobHasenbusch:physicaa1993}%
  \BibitemOpen
  \bibfield  {author} {\bibinfo {author} {\bibfnamefont {A.~P.}\ \bibnamefont
  {Gottlob}}\ and\ \bibinfo {author} {\bibfnamefont {M.}~\bibnamefont
  {Hasenbusch}},\ }\href {\doibase 10.1016/0378-4371(93)90131-M} {\bibfield
  {journal} {\bibinfo  {journal} {Physica A}\ }\textbf {\bibinfo {volume}
  {201}},\ \bibinfo {pages} {593} (\bibinfo {year} {1993})}\BibitemShut
  {NoStop}%
\bibitem [{Note6()}]{Note6}%
  \BibitemOpen
  \bibinfo {note} {These configurations can be generalized to the case in which
  there is a different constant angle for each of the $d$ independent
  directions of the lattice; however, for simplicity we shall restrict to the
  case of just one angle, equal for all the directions.}\BibitemShut {Stop}%
\bibitem [{\citenamefont {Demazure}(2000)}]{Demazure}%
  \BibitemOpen
  \bibfield  {author} {\bibinfo {author} {\bibfnamefont {M.}~\bibnamefont
  {Demazure}},\ }\href@noop {} {\emph {\bibinfo {title} {Bifurcations and
  Catastrophes: {G}eometry of Solutions to Nonlinear Problems}}}\ (\bibinfo
  {publisher} {Springer, Berlin},\ \bibinfo {year} {2000})\BibitemShut
  {NoStop}%
\bibitem [{\citenamefont {von Smekal}\ \emph {et~al.}(2007)\citenamefont {von
  Smekal}, \citenamefont {Mehta}, \citenamefont {Sternbeck},\ and\
  \citenamefont {Williams}}]{vonSmekal:2007ns}%
  \BibitemOpen
  \bibfield  {author} {\bibinfo {author} {\bibfnamefont {L.}~\bibnamefont {von
  Smekal}}, \bibinfo {author} {\bibfnamefont {D.}~\bibnamefont {Mehta}},
  \bibinfo {author} {\bibfnamefont {A.}~\bibnamefont {Sternbeck}}, \ and\
  \bibinfo {author} {\bibfnamefont {A.~G.}\ \bibnamefont {Williams}},\
  }\href@noop {} {\bibfield  {journal} {\bibinfo  {journal} {PoS}\ }\textbf
  {\bibinfo {volume} {LAT2007}},\ \bibinfo {pages} {382} (\bibinfo {year}
  {2007})}\BibitemShut {NoStop}%
\bibitem [{\citenamefont {Bates}\ \emph {et~al.}()\citenamefont {Bates},
  \citenamefont {Hauenstein}, \citenamefont {Sommese},\ and\ \citenamefont
  {Wampler}}]{BHSW06}%
  \BibitemOpen
  \bibfield  {author} {\bibinfo {author} {\bibfnamefont {D.~J.}\ \bibnamefont
  {Bates}}, \bibinfo {author} {\bibfnamefont {J.~D.}\ \bibnamefont
  {Hauenstein}}, \bibinfo {author} {\bibfnamefont {A.~J.}\ \bibnamefont
  {Sommese}}, \ and\ \bibinfo {author} {\bibfnamefont {C.~W.}\ \bibnamefont
  {Wampler}},\ }\href@noop {} {\enquote {\bibinfo {title} {Bertini: Software
  for numerical algebraic geometry},}\ }\bibinfo {howpublished}
  {\url{http://www.nd.edu/~sommese/bertini}}\BibitemShut {NoStop}%
\bibitem [{\citenamefont {Allgower}\ and\ \citenamefont
  {Georg}(1979)}]{79:allgower}%
  \BibitemOpen
  \bibfield  {author} {\bibinfo {author} {\bibfnamefont {E.~L.}\ \bibnamefont
  {Allgower}}\ and\ \bibinfo {author} {\bibfnamefont {K.}~\bibnamefont
  {Georg}},\ }\href@noop {} {\emph {\bibinfo {title} {Introduction to Numerical
  Continuation Methods}}}\ (\bibinfo  {publisher} {John Wiley \& Sons, New
  York},\ \bibinfo {year} {1979})\BibitemShut {NoStop}%
\bibitem [{\citenamefont {Mehta}(2011{\natexlab{b}})}]{Mehta:2011wj}%
  \BibitemOpen
  \bibfield  {author} {\bibinfo {author} {\bibfnamefont {D.}~\bibnamefont
  {Mehta}},\ }\href {\doibase 10.1155/2011/263937} {\bibfield  {journal}
  {\bibinfo  {journal} {Adv. High Energy Phys.}\ }\textbf {\bibinfo {volume}
  {2011}},\ \bibinfo {pages} {263937} (\bibinfo {year}
  {2011}{\natexlab{b}})}\BibitemShut {NoStop}%
\bibitem [{\citenamefont {Maniatis}\ and\ \citenamefont
  {Mehta}(2012)}]{Maniatis:2012ex}%
  \BibitemOpen
  \bibfield  {author} {\bibinfo {author} {\bibfnamefont {M.}~\bibnamefont
  {Maniatis}}\ and\ \bibinfo {author} {\bibfnamefont {D.}~\bibnamefont
  {Mehta}},\ }\href {\doibase 10.1140/epjp/i2012-12091-1} {\bibfield  {journal}
  {\bibinfo  {journal} {Eur. Phys. J. Plus}\ }\textbf {\bibinfo {volume}
  {127}},\ \bibinfo {pages} {91} (\bibinfo {year} {2012})}\BibitemShut
  {NoStop}%
\bibitem [{\citenamefont {Sommese}\ \emph {et~al.}(2005)\citenamefont
  {Sommese}, \citenamefont {Verschelde},\ and\ \citenamefont
  {Wampler}}]{SVW:96}%
  \BibitemOpen
  \bibfield  {author} {\bibinfo {author} {\bibfnamefont {A.~J.}\ \bibnamefont
  {Sommese}}, \bibinfo {author} {\bibfnamefont {J.}~\bibnamefont {Verschelde}},
  \ and\ \bibinfo {author} {\bibfnamefont {C.~W.}\ \bibnamefont {Wampler}},\
  }in\ \href@noop {} {\emph {\bibinfo {booktitle} {Solving Polynomial
  Equations: Foundations, Algorithms, and Applications}}},\ \bibinfo {series}
  {Algorithms and Computation in Mathematics}, Vol.~\bibinfo {volume} {14},\
  \bibinfo {editor} {edited by\ \bibinfo {editor} {\bibfnamefont
  {A.}~\bibnamefont {Dickenstein}}\ and\ \bibinfo {editor} {\bibfnamefont
  {I.~Z.}\ \bibnamefont {Emiris}}}\ (\bibinfo  {publisher} {Springer},\
  \bibinfo {address} {Berlin},\ \bibinfo {year} {2005})\ pp.\ \bibinfo {pages}
  {339--392}\BibitemShut {NoStop}%
\bibitem [{\citenamefont {Mehta}\ \emph
  {et~al.}(2012{\natexlab{a}})\citenamefont {Mehta}, \citenamefont {He},\ and\
  \citenamefont {Hauenstein}}]{Mehta:2012wk}%
  \BibitemOpen
  \bibfield  {author} {\bibinfo {author} {\bibfnamefont {D.}~\bibnamefont
  {Mehta}}, \bibinfo {author} {\bibfnamefont {Y.-H.}\ \bibnamefont {He}}, \
  and\ \bibinfo {author} {\bibfnamefont {J.~D.}\ \bibnamefont {Hauenstein}},\
  }\href {\doibase 10.1007/JHEP07(2012)018} {\bibfield  {journal} {\bibinfo
  {journal} {JHEP}\ }\textbf {\bibinfo {volume} {1207}},\ \bibinfo {pages}
  {018} (\bibinfo {year} {2012}{\natexlab{a}})}\BibitemShut {NoStop}%
\bibitem [{\citenamefont {Lu}\ \emph {et~al.}(2006)\citenamefont {Lu},
  \citenamefont {Bates}, \citenamefont {Sommese},\ and\ \citenamefont
  {Wampler}}]{Lu06findingall}%
  \BibitemOpen
  \bibfield  {author} {\bibinfo {author} {\bibfnamefont {Y.}~\bibnamefont
  {Lu}}, \bibinfo {author} {\bibfnamefont {D.~J.}\ \bibnamefont {Bates}},
  \bibinfo {author} {\bibfnamefont {A.~J.}\ \bibnamefont {Sommese}}, \ and\
  \bibinfo {author} {\bibfnamefont {C.~W.}\ \bibnamefont {Wampler}},\ }in\
  \href@noop {} {\emph {\bibinfo {booktitle} {Algebra, Geometry and Their
  Interactions}}},\ \bibinfo {series} {Contemporary Mathematics}, Vol.\
  \bibinfo {volume} {448},\ \bibinfo {editor} {edited by\ \bibinfo {editor}
  {\bibfnamefont {A.}~\bibnamefont {Corso}}, \bibinfo {editor} {\bibfnamefont
  {J.}~\bibnamefont {Migliore}}, \ and\ \bibinfo {editor} {\bibfnamefont
  {C.}~\bibnamefont {Polini}}}\ (\bibinfo  {publisher} {American Mathematical
  Society},\ \bibinfo {year} {2006})\ pp.\ \bibinfo {pages}
  {183--205}\BibitemShut {NoStop}%
\bibitem [{\citenamefont {Mehta}\ \emph
  {et~al.}(2012{\natexlab{b}})\citenamefont {Mehta}, \citenamefont
  {Hauenstein},\ and\ \citenamefont {Kastner}}]{MehtaHauensteinKastner12}%
  \BibitemOpen
  \bibfield  {author} {\bibinfo {author} {\bibfnamefont {D.}~\bibnamefont
  {Mehta}}, \bibinfo {author} {\bibfnamefont {J.~D.}\ \bibnamefont
  {Hauenstein}}, \ and\ \bibinfo {author} {\bibfnamefont {M.}~\bibnamefont
  {Kastner}},\ }\href@noop {} {\bibfield  {journal} {\bibinfo  {journal} {Phys.
  Rev. E}\ }\textbf {\bibinfo {volume} {85}},\ \bibinfo {pages} {061103}
  (\bibinfo {year} {2012}{\natexlab{b}})}\BibitemShut {NoStop}%
\bibitem [{\citenamefont {Strogatz}\ and\ \citenamefont
  {Mirollo}(1991)}]{strogatz1991stability}%
  \BibitemOpen
  \bibfield  {author} {\bibinfo {author} {\bibfnamefont {S.}~\bibnamefont
  {Strogatz}}\ and\ \bibinfo {author} {\bibfnamefont {R.}~\bibnamefont
  {Mirollo}},\ }\href@noop {} {\bibfield  {journal} {\bibinfo  {journal}
  {Journal of Statistical Physics}\ }\textbf {\bibinfo {volume} {63}},\
  \bibinfo {pages} {613} (\bibinfo {year} {1991})}\BibitemShut {NoStop}%
\bibitem [{\citenamefont {Griffiths}(1972)}]{Griffiths:inDombGreen}%
  \BibitemOpen
  \bibfield  {author} {\bibinfo {author} {\bibfnamefont {R.~B.}\ \bibnamefont
  {Griffiths}},\ }in\ \href@noop {} {\emph {\bibinfo {booktitle} {Phase
  Transitions and Critical Phenomena}}},\ Vol.~\bibinfo {volume} {1},\ \bibinfo
  {editor} {edited by\ \bibinfo {editor} {\bibfnamefont {C.}~\bibnamefont
  {Domb}}\ and\ \bibinfo {editor} {\bibfnamefont {M.~S.}\ \bibnamefont
  {Green}}}\ (\bibinfo  {publisher} {Academic Press},\ \bibinfo {address}
  {London},\ \bibinfo {year} {1972})\BibitemShut {NoStop}%
\bibitem [{\citenamefont {Franzosi}\ and\ \citenamefont
  {Pettini}(2004)}]{FranzosiPettini:prl}%
  \BibitemOpen
  \bibfield  {author} {\bibinfo {author} {\bibfnamefont {R.}~\bibnamefont
  {Franzosi}}\ and\ \bibinfo {author} {\bibfnamefont {M.}~\bibnamefont
  {Pettini}},\ }\href {\doibase 10.1103/PhysRevLett.92.060601} {\bibfield
  {journal} {\bibinfo  {journal} {Phys. Rev. Lett.}\ }\textbf {\bibinfo
  {volume} {92}},\ \bibinfo {pages} {060601} (\bibinfo {year}
  {2004})}\BibitemShut {NoStop}%
\bibitem [{\citenamefont {Franzosi}\ and\ \citenamefont
  {Pettini}(2007)}]{FranzosiPettini:npb}%
  \BibitemOpen
  \bibfield  {author} {\bibinfo {author} {\bibfnamefont {R.}~\bibnamefont
  {Franzosi}}\ and\ \bibinfo {author} {\bibfnamefont {M.}~\bibnamefont
  {Pettini}},\ }\href {\doibase DOI: 10.1016/j.nuclphysb.2007.04.035}
  {\bibfield  {journal} {\bibinfo  {journal} {Nucl. Phys. B}\ }\textbf
  {\bibinfo {volume} {782}},\ \bibinfo {pages} {219} (\bibinfo {year}
  {2007})}\BibitemShut {NoStop}%
\end{thebibliography}%

\end{document}